\documentclass[twocolumn,superscriptaddress,floatfix,preprintnumbers, nofootinbib,hyperref,floatfix]{revtex4-2} 
\pdfoutput=1
\usepackage[colorlinks=true,breaklinks=true]{hyperref}
\usepackage[normalem]{ulem}
\usepackage[utf8]{inputenc}
\hypersetup{allcolors=[rgb]{0.0 0.0 0.6},linkcolor=[rgb]{0.75 0.05 0.05}}
\usepackage{amsmath,amssymb}
\usepackage{epsfig}  
\usepackage{graphicx}   
\usepackage{slashed}  

\usepackage{empheq,etoolbox}
\patchcmd{\subequations}
  {\theparentequation\alph{equation}}
  {\theparentequation.\alph{equation}}
  {}{}

\usepackage{url}
\usepackage{color}
\usepackage{multirow}
\usepackage{subfigure}
\usepackage{comment}
\usepackage[dvipsnames]{xcolor}
\usepackage{letltxmacro}
\LetLtxMacro{\oldcite}{\cite}
\renewcommand{\cite}[1]{\mbox{\oldcite{#1}}}

\DeclareMathOperator{\cm}{cm}

\DeclareMathOperator{\MeV}{MeV}

\DeclareMathOperator{\s}{s}

\newcommand{\beq}{\begin{equation}}
\newcommand{\eeq}{\end{equation}}

\allowdisplaybreaks

\bibpunct{[}{]}{,}{n}{}{,}

\definecolor{darkblue}{rgb}{1, 0.1, 0.2}

\begin{document}

\title{511 keV line constraints on feebly interacting particles from supernovae}

\author{Francesca Calore}
\email{calore@lapth.cnrs.fr}
\affiliation{LAPTh, USMB, CNRS,  F-74940 Annecy, France}

\author{Pierluca Carenza}
\email{pierluca.carenza@fysik.su.se}
\affiliation{The Oskar Klein Centre, Department of Physics, Stockholm University, Stockholm 106 91, Sweden}
\affiliation{Dipartimento Interateneo di Fisica ``Michelangelo Merlin'', Via Amendola 173, 70126 Bari, Italy.}
\affiliation{Istituto Nazionale di Fisica Nucleare - Sezione di Bari,
Via Orabona 4, 70126 Bari, Italy.}

\author{Maurizio Giannotti}
\email{MGiannotti@barry.edu}
\affiliation{Department of Chemistry and Physics, Barry University, 11300 NE 2nd Ave., Miami Shores, FL 33161, USA}

\author{Joerg Jaeckel}
\email{jjaeckel@thphys.uni-heidelberg.de}
\affiliation{Institut f\"ur theoretische Physik, Universit\"at Heidelberg,
Philosophenweg 16, 69120 Heidelberg, Germany}

\author{Giuseppe Lucente}
\email{giuseppe.lucente@ba.infn.it }
\affiliation{Dipartimento Interateneo di Fisica ``Michelangelo Merlin'', Via Amendola 173, 70126 Bari, Italy.}
\affiliation{Istituto Nazionale di Fisica Nucleare - Sezione di Bari,
Via Orabona 4, 70126 Bari, Italy.}

\author{Leonardo Mastrototaro}
\email{lmastrototaro@unisa.it}
\affiliation{Dipartimento di Fisica ``E.R Caianiello'', Università degli Studi di Salerno,
Via Giovanni Paolo II, 132 - 84084 Fisciano (SA), Italy.}
\affiliation{Istituto Nazionale di Fisica Nucleare - Gruppo Collegato di Salerno - Sezione di Napoli, Via Giovanni Paolo II, 132 - 84084 Fisciano (SA), Italy.}

\author{Alessandro Mirizzi}
\email{alessandro.mirizzi@ba.infn.it }
\affiliation{Dipartimento Interateneo di Fisica ``Michelangelo Merlin'', Via Amendola 173, 70126 Bari, Italy.}
\affiliation{Istituto Nazionale di Fisica Nucleare - Sezione di Bari,
Via Orabona 4, 70126 Bari, Italy.}


\begin{abstract}
Feebly interacting particles with masses
with $\mathcal O$(10-100) MeV can be copiously produced by core-collapse supernovae (SNe). In this paper we consider the case of MeV-ish sterile neutrinos and dark photons \emph{mixed} with ordinary neutrinos and photons, respectively. Furthermore, both sterile neutrinos and dark photons may decay into positrons on their route to Earth. Such positrons would  annihilate with electrons in the Galactic medium and contribute to the photon flux in the 511 keV line. Using the SPI (SPectrometer on INTEGRAL) observation of this line improves the bounds on the mixing parameters for these particles by several orders of magnitude below what is already excluded by the SN 1987A energy-loss argument.
 \end{abstract}

\maketitle

\section{Introduction}
Core-collapse supernovae (SNe) are  powerful cosmic sources of light weakly interacting particles. Notably, during a  SN explosion, the amazing number ${\mathcal O} (10^{58})$ of (anti)neutrinos of all flavours are emitted with average energies $E \sim 15$~MeV. These can be used to probe fundamental properties of neutrinos, e.g.~mixing and nonstandard interactions  (see, Ref.~\cite{Raffelt:1996wa} for a pedagogical introduction and, e.g.~\cite{Mirizzi:2015eza,Horiuchi:2018ofe}, for more recent works). More generally, also new feebly interacting particles (FIPs)~\cite{Agrawal:2021dbo} can be abundantly produced in a SN core. For typical core temperatures $T \simeq \mathcal{O}(30)$~MeV, particles with masses up to $\mathcal{O}(100)$~MeV can be emitted without being Boltzmann suppressed. If free streaming, these particles would constitute a novel channel of \emph{energy loss}, shortening the duration of the neutrino burst~\cite{Raffelt:2012kt,Chang:2018rso}. Notably, a variety of new physics scenarios has been constrained using this powerful argument in relation to the SN 1987A neutrino detection~\cite {Raffelt:1987yt}. An incomplete list of novel particles include axions~\cite{Raffelt:1987yt,Keil:1996ju,Chang:2018rso,Carenza:2019pxu,Carenza:2020cis}, scalar bosons~\cite{Caputo:2021rux}, sterile neutrinos~\cite{Kolb:1996pa,Raffelt:2011nc,Mastrototaro:2019vug}, dark photons~\cite{Chang:2016ntp}, light $CP$-even scalars~\cite{Dev:2020eam}, dark flavored particles~\cite{Camalich:2020wac} and unparticles~\cite{Hannestad:2007ys} (see~\cite{Raffelt:2011zab} for a list of reviews on this topic and for additional references).

The next detection of a high-statistics SN neutrino burst from a Galactic event in a large underground detector is  expected to significantly improve the previous bounds based on the sparse SN 1987A neutrino data (see, e.g.,~\cite{Fischer:2016cyd,Fischer:2021jfm} for the case of axions). Therefore, such an event can be considered as one of the next frontiers of low-energy neutrino and FIP astronomy.

Furthermore, aside from the \emph{indirect} signature imprinted  on the neutrino burst by an exotic energy loss, one can also search  for a \emph{direct} signature of novel particles emission from SNe. Notably in the case of ultralight axion-like particles (ALPs) with $m_a \lesssim 10^{-11}$~eV, coupled with photons, their conversions into gamma rays in the Galactic magnetic field offer a clear signature. Following this idea, dedicated searches for an ALP induced gamma-ray signal have been performed in relation to SN 1987A (see, e.g.,~\cite{Payez:2014xsa}), in case of future Galactic SN explosions~\cite{Meyer:2016wrm}, for extra-galactic SNe~\cite{Meyer:2020vzy,Crnogorcevic:2021wyj}, and in relation to the observation of the diffuse gamma-ray background~\cite{Calore:2020tjw,Calore:2021hhn}.

Heavy ALPs (with $m_a \sim 1-10$~MeV) coupled with photons would decay on their path to Earth leading to another peculiar gamma-ray flux~\cite{Giannotti:2010ty,Jaeckel:2017tud}. Some of us recently studied also the case of heavy ALPs coupled with nucleons and electrons~\cite{Calore:2021klc}. These ALPs would be efficiently produced in a SN core via the nucleon-nucleon bremmstrahlung~\cite{Carenza:2019pxu}. Then, for masses $m_a\gtrsim 1$~MeV, they would decay into electron-positron pairs, generating a positron flux. In the case of Galactic SNe, the annihilation of the created positrons with the electrons present in the Galaxy would contribute to the 511 keV annihilation line. Using the SPI (SPectrometer on INTEGRAL)~\cite{Strong:2005zx,Bouchet:2010dj,Siegert:2015knp, Siegert:2019tus} observation of this line allows one to exclude a wide range of the axion-electron coupling.

The goal of our present paper is to extend the 511 keV photon constraints on ALPs to sterile neutrinos and dark photons (see, e.g.,~\cite{Agrawal:2021dbo,Alekhin:2015byh,Appelquist:2002me,Asaka:2005an,Abel:2008ai,Jaeckel:2010ni} for motivations for these particles and further literature). These FIPs share two important features: 1) They can be produced via \emph{mixing} with Standard Model (SM) particles (the SM neutrino and the photon, respectively); 2) They have decay channels with positrons in the final state. While the former is just a theoretical similarity, the latter is a fundamental aspect of our analysis and leads to the signal that we focus on: the positrons are produced as decay products, they are slowed down and eventually annihilate, sourcing 511 keV photons.

The 511 keV photon bound  was considered in the seminal paper~\cite{Dar:1986wb} (see also Sec.~12.5.1 of Ref.~\cite{Raffelt:1996wa}) for the case of decaying heavy neutrinos $\nu_H$, using the early measurements of the 511 keV gamma-ray flux~\cite{osti_7251482}. For dark photons the 511 keV bound was recently studied in~Ref.~\cite{DeRocco:2019njg}. In the present paper we improve  on these previous  results by using the limits on the positron injection obtained in~\cite {Calore:2021klc} that account for the spatial morphology of the signal. 
Moreover, we also consider a broader range of SN models (as in Ref.~\cite{Calore:2021hhn} for the case of ALPs coupled to photons).

The plan for this paper is as follows. In Sec.~\ref{sec:positrons} we recall the signal from positrons produced in SN by a decaying massive particle. Then in Sec.~\ref{sec:production_SN} and Sec.~\ref{sec:darkphotons} we discuss the production and decay as well as the resulting new limits for sterile neutrinos and dark photons, respectively. We wrap up the discussion in Sec.~\ref{sec:conclusions}. In Appendices \ref{app:sterile} and \ref{app:dpprod} we discuss the details of the sterile neutrino and dark photon production, respectively. Finally, in Appendix \ref{app:uncertainties} we present in detail the uncertainties affecting our bounds.

\section{Bounds on FIPs from the 511 keV line observation}
\label{sec:positrons}

Any FIP $X$ that is sufficiently weakly coupled to SM fields, once produced in the SN core can escape without  interacting with the stellar medium. In many cases there is a large portion of the  parameter space where this process allows a very efficient production of such FIPs. The energy-loss bound from SN 1987A requires that the total luminosity carried by  an exotic species should be~\cite{Raffelt:1990yz,Caputo:2021rux} 
\begin{equation}
L_X^{\rm tot} \lesssim 3 \times 10^{52} \,\ \textrm{erg} \,\ \textrm{s}^{-1} \,\ ,
\label{eq:lumenX}
\end{equation}
at the beginning of the cooling phase (typically at $t_{\rm pb}\simeq1$~s, where $t_{\rm pb}$ is the post-bounce time). Furthermore, for electromagnetic FIP decays, like the ones we consider here, a possible improvement of the energy-loss bound was proposed in Ref.~\cite{Sung:2019xie} requiring that the energy transfer to the outer layers from FIP decay products does not lead to too energetic SN explosions. In case of an efficient energy transfer one can exclude FIPs with an emissivity $\sim 2$ order of magnitudes below what probed by the energy-loss limit, i.e.~a luminosity going into an electromagnetic channel
\begin{equation}
L_X^{\rm e.m.} \lesssim 2 \times 10^{50} \,\ \textrm{erg} \,\ \textrm{s}^{-1} \,\ . 
\label{eq:lumenX_1}
\end{equation}
However, for particles decaying outside the SN envelope this latter argument does not apply.

\begin{figure*}[t!]
\vspace{0.cm}
\includegraphics[width=1.6\columnwidth]{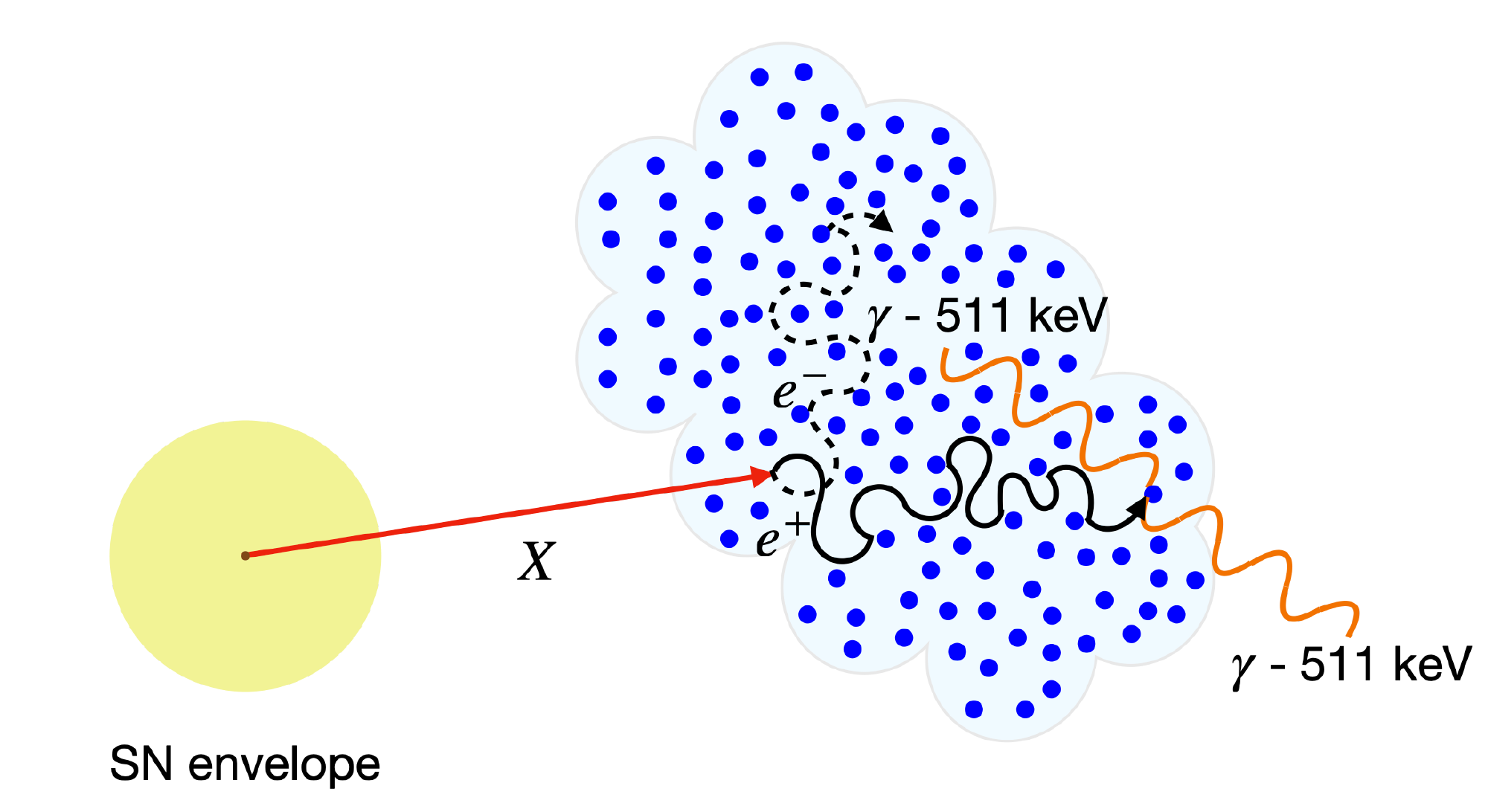}
\caption{A schematic representation of the production mechanism of the 511 keV gamma-ray line by $X$ decays in electron-positron pairs. The $e^+$ and $e^-$ trajectories are not straight due to the presence of an inhomogeneous magnetic field. They are also slowed down due to scattering.}
\label{fig:sketch}
\end{figure*}

Besides these constraints, an emissivity close to the luminosity bound in Eq.~(\ref{eq:lumenX_1}) would lead to an enormous flux. Its possible detection, and the analysis of observable signatures associated with it, is subject of a number of studies (cf. e.g. Refs.~\cite{Dar:1986wb,DeRocco:2019njg,Calore:2021klc}) including the present one. If the particle under examination is allowed to decay into electron-positron pairs outside the SN envelope, one important observational consequence is a contribution to the 511 keV photon line, measured by the  SPI~\cite{Strong:2005zx,Bouchet:2010dj,Siegert:2015knp, Siegert:2019tus}. 
The mechanism proceeds as follows. As schematically depicted in Fig.~\ref{fig:sketch}, the FIPs $X$ under consideration are allowed to decay into electron-positron pairs outside the stellar matter, i.e.
\begin{equation}
X \to n_{i}e^+ + Y_{i} \,\ .    
\end{equation}
Here, $n_i$ is the number of positrons produced in the $i$-th process and $Y_i$ other species produced in the decays.
The resulting positrons are efficiently slowed down through elastic Bhabha scatterings with the electrons in the Galaxy (the blue dots in Fig.~\ref{fig:sketch}) and annihilate, almost at rest, after the formation of positronium bound states~\cite{Guessoum:2005cb}. Approximately $25\%$ of them form (singlet) para-positronium states, which eventually decay into two photons with opposite momentum and total energy almost exactly equal to $2m_e$.\footnote{Annihilation through the (triplet) orthopositronium state, formed in the $75\%$ of cases, leads to three photons that do not contribute to the 511~keV photon line.}

Below, we provide some general results about this process closely following~\cite{Calore:2021klc}, to which we address the interested reader for further details. In the next Sections, we will focus on the specific examples of sterile neutrinos and dark photons.
 
Considering the contribution of the Galactic population of SNe II and SNe Ib/c, in complete generality,\footnote{We remark that the contribution of SNe Ib/c was neglected for simplicity in Ref.~\cite{Calore:2021klc}.} the average number of positrons produced in FIP decays outside the SN photosphere is
\begin{eqnarray}
    N_{\rm pos}\!\!& = &\!\!n_{\rm pos,X}\int dE\frac{dN_{X}^{0}}{dE}\left(\epsilon_{II}e^{-{r_{II}}/{l_{X}}}+\epsilon_{I}\,e^{-{r_{\rm I}}/{l_{X}}}\right) \nonumber
    \\
    &\times&\bigg[1 - \exp\bigg(-\frac{r_{\rm G}}{l_{X}}\bigg) \bigg] \;,
    \label{eq:npos}
\end{eqnarray}
where the FIP production rate $dN_{X}^{0}/dE$ depends on the studied particle and it will be specified later. We indicate with $l_{X}$ the total FIP decay length, and with 
\begin{equation}
\label{eq:averagepositrons}
    n_{\rm pos,X}=\sum_{i}n_i BR(X\to n_ie^{+}+Y_{i}),
\end{equation}
the average number of positrons produced in a FIP decay. The quantity $r_{\rm G}$ in Eq.~(\ref{eq:npos}) is a typical escape radius from the Galaxy. In the following we conservatively assume it to be $1$~kpc, in order to stay inside the Galaxy in all directions, in particular the one perpendicular to the Galactic plane. Moreover, following~\cite{DeRocco:2019njg} we fix
\begin{equation}
    r_{II}=10^{14}~{\rm cm}, \qquad  r_{I}=2\times 10^{12}~{\rm cm}\,,
\end{equation}
for the envelope radii of type II and Ib/c SNe, while according to Ref.~\cite{Li:2010kd}, we take as average fractions of SNe of type II and Ib/c 
\begin{equation}
    \epsilon_{II}=1-\epsilon_{I},\qquad \epsilon_{I}=0.33\,\ .
\end{equation}

From Eq.~(\ref{eq:npos}), one can predict the main qualitative features in the behaviour of the positrons produced by FIP decays. Assuming that a FIP $X$ is produced through a coupling $g_{X,P}$ and decays due to a coupling $g_{X,D}$, then, $dN_{X}^{0}/dE\propto g_{X,P}^{2}$ and $l_{X}\propto g_{X,D}^{-2}$. Therefore, from Eq.~(\ref{eq:npos}), one can easily predict the following limiting regimes: 
\begin{subequations}\label{cases}
\begin{empheq}[left={N_{\rm pos} \propto   \empheqlbrace\,}]{align}
 &g_{X,P}^{2}\,g_{X,D}^{2}& \,  l_{X}\gg r_{\rm G} \,\ , \label{rg}\\
  &g_{X,P}^{2}&       r_{II}\ll l_{X}\ll r_{\rm G} \,\  , \label{rii}\\
&g_{X,P}^{2}\,\exp\left(-g_{X,D}^{2}\right)&         l_{X}\ll r_{I} \,\   . \label{ri} 
\end{empheq}
\end{subequations}

For $r_{I} < l_{X} < r_{II}$ the behaviour is non-trivial since the contributions of SNe Ib/c and SNe II are both relevant. In the case under consideration, we will always find $l_X<r_G$. Therefore, only Eqs.~(\ref{rii}) and (\ref{ri}) will be relevant in this work.

For our numerical estimations, we will assume as a benchmark the SN simulations of Ref.~\cite{Fischer:2018kdt}, based on the
general relativistic neutrino radiation hydrodynamics
model AGILE-BOLTZTRAN, featuring three-flavor Boltzmann neutrino transport~\cite{Mezzacappa:1993gn,Liebendoerfer:2002xn} and including a complete
set of weak interactions (see Table I in Ref.~\cite{Fischer:2018kdt}). The SN simulations were launched
from the $18~M_\odot$ progenitor of the stellar evolution calculations of Ref.~\cite{Woosley:2002zz}. We choose this particular progenitor mass, since among the models at our disposal it is the one closest to the average successful SN explosion progenitor mass
\begin{equation}
    \langle M \rangle = \frac{\int_{8~M_\odot}^{50~M_\odot} dM M \phi(M)}{\int_{8~M_\odot}^{50~M_\odot} dM \phi(M)}\approx 16~M_\odot\,,
\end{equation}
where $\phi(M)\sim M^{-2.35}$ is the modified Salpeter-A initial mass function of Ref.~\cite{Baldry:2003xi}. We note, however, the variance is quite large and approximately equal to $\sim 9~M_{\odot}$. Therefore, we generally expect SNe with different progenitor masses to contribute. As discussed in Ref.~\cite{Calore:2021klc} the typical lifetime of the positrons is of the order of $\tau_{e}\sim 10^{3}-10^{6}~{\rm years}$ and a single SN is expected to contribute to the signal for a similar amount of time. Accordingly, at any given time $\sim 10-10^4$ SNe may be active in the signal. In particular at the lower end of this range fluctuations may be sizeable.  In light of this uncertainty and for simplicity we consider the whole signal being generated by a single exemplary progenitor mass. The impact of the chosen progenitor mass on the derived bounds is then discussed in more detail in Appendix \ref{app:uncertainties}, where the 511 keV photon constraint is re-evaluated using two other SN models, with progenitor mass $8.8~M_\odot$ and $25~M_\odot$ respectively, reported in Ref.~\cite{Fischer:2009af}.

As discussed in \cite{Calore:2021klc}, the photon line signal at 511 keV is
\begin{equation}
\frac{d \phi_\gamma^{ 511}}{d\Omega}= 2k_{ps}N_{\rm pos} \Gamma_{cc}\int ds \,  s^2 \frac{n_{cc}[r(s,b,l),z(s,b)]}{4 \pi s^2} \;,
\label{eq:phot}
\end{equation}
where $d\Omega = dl\, db \cos b$, with $- \pi \leq l \leq \pi$ being the longitude and $-\pi/2 \leq b \leq \pi/2$ being the latitude in the Galactic coordinate system $(s,b,l)$, with $s$ as the distance from the SN to the Sun. Moreover, $k_{ps}=1/4$ accounts for the fraction of positrons annihilating through para-positronium,\footnote{We recall that annihilation through orthopositronium leads to three photons that do not contribute to the 511~keV photon line.} then producing two photons with energy equal to 511~keV~\cite{Karshenboim:2003vs}. According to Ref.~\cite{Li:2010kd}, we fix $\Gamma_{cc}=2.30$ SNe/century as the Galactic SN rate. Finally, $n_{cc}$ is the SN volume distribution~\cite{Mirizzi:2006xx} in the Galactocentric coordinate system $(r,z,l)$, with $r$ the galactocentric radius and $z$ the height above the Galactic plane, connected with the Galactic coordinate system through the relations
\begin{eqnarray}
&& r = \sqrt{s^2 \cos^2 b + d_{\odot}^2 -2 d_\odot s\cos l \cos b}\,\ , \\ 
&& z=s \sin b \,,
\end{eqnarray}
where $d_{\odot}=8.5$ kpc is the solar distance from the Galactic center. In order to account for the distance travelled by positrons before being stopped, we smear the SN distribution over the positron propagation scale $\lambda$ in the Galaxy which we fix at $\lambda=1$~kpc, based on typical injection energies and interstellar medium conditions, see Ref.~\cite{Calore:2021klc} and~\cite{Jean:2005af,Jean:2009zj,Martin:2012hv} for details of this choice, and Appendix~\ref{app:uncertainties} for a discussion of the impact on the resulting limits. From there we see that long distance propagation of the positrons can lead to a noticeable weakening of the limits. This highlights the importance of improving the measurements and modelling of this process (see Ref.~\cite{Siegert:2021trw} for first steps in this direction).

A bound can be placed by requiring that the flux predicted in Eq.~\eqref{eq:phot} is below the $2\sigma$ confidence maximal value measured by SPI~\cite{Strong:2005zx,Bouchet:2010dj,Siegert:2015knp, Siegert:2019tus}. We will use observations of the 511 keV line from~\cite{Siegert:2019tus} to constrain the positron flux injected by FIPs produced by Galactic SNe and decaying outside the SN envelope. Precisely, given the shape of the photon signal, the most constraining bin is the one with longitude $l\in [28.25^{\circ};31.25^{\circ}]$ and latitude $b\in [-10.75^{\circ};10.25^{\circ}]$, as discussed in Ref.~\cite{Calore:2021klc}. Therefore the condition that we will require in the following is
\begin{equation}
    \int d\Omega \frac{d \phi_\gamma^{ 511}}{d\Omega}\lesssim 8.35\times10^{-6} \cm^{-2}\s^{-1}\;,
\end{equation}
where the integral is performed over the $(l,b)$ range indicated above. This condition implies a bound on the number of positrons injected in the Galaxy\footnote{As in Ref.~\cite{Calore:2021klc},  this number accounts also for the positrons annihilating in flight by conservatively assuming that this fraction is at most $25\%$ in the relevant energy range~\cite{Beacom:2005qv}. We stress that the present bound on $N_{\rm pos}$ is different from the one quoted in Ref.~\cite{Calore:2021klc}, i.e.~$N_{\rm pos}\lesssim 1.6\times 10^{52}$, since in that work we used $\Gamma_{cc}=2$ SNe/century.}
\begin{equation}
N_{\rm{pos}}\lesssim 1.4\times 10^{52} \,\ .
\label{eq:nposbound}
\end{equation}
In the following, we will apply this analysis to the cases of sterile neutrinos and dark photons.

We remark that, in the present work, we only consider the contribution to the 511 keV line signal, neglecting the additional constraining power that may come from the full computation of the diffuse gamma-ray emission induced by FIPs in the SPI energy band. This task would require us to keep track of the full energy dependence of injected positrons and electrons and perform a dedicated spectral analysis. As such, this is beyond the scope of the present work, and we postpone this full treatment to a future work.

\section{Sterile neutrinos}
\label{sec:production_SN}

\subsection{Positron flux}
\label{sec:sterposflux}

We consider heavy sterile neutrinos with masses 10 MeV $\lesssim m_s\lesssim $ 200 MeV~\cite{Asaka:2005an,Asaka:2005pn}, mixed dominantly with one active neutrino $\nu_\alpha$ ($\alpha=e,\mu,\tau$) as
\begin{eqnarray}
\nu_\alpha &=& \cos \theta_{\alpha s} \nu_\ell + \sin\theta_{\alpha s} \nu_H \,\ ,  \nonumber \\ 
\nu_s &=& -\sin \theta_{\alpha s} \nu_\ell + \cos\theta_{\alpha s} \nu_H \,\ ,
\end{eqnarray}
where $\nu_\ell$ and $\nu_H$ are a light and a heavy mass eigenstate, respectively, and $\theta_{\alpha s} \ll 1$, i.e.  $\nu_\ell$ is mostly active and $\nu_H$ is mostly sterile. We can relate the  mixing angle to the unitary mixing matrix $U$, where
\begin{equation}
|U_{\alpha s}|^2 \simeq \frac{1}{4} \sin^2 2 \theta_{\alpha s}\simeq \theta_{\alpha s}^2 \,\ .
\end{equation}
Through neutral-current interactions, $\nu_H$ can decay into a $\nu_\ell$ and a pair of other light leptons. It can also scatter with other species in the plasma.

\begin{table}[t]
    \centering
    \begin{tabular}{|c|c|c|}
    \hline
    Process & $\Gamma/G_F^2m_s^3|U_{\tau s}|^2$ & Threshold (MeV)\\
    \hline
    $\nu_s\rightarrow\nu_\tau\gamma$& $9\alpha m_s^2/2048\pi^4$ & $0$\\
    $\nu_s\rightarrow\nu_\tau\nu_\tau\bar{\nu}_\tau$& $m_s^2/384\pi^3$ & $0$\\
    $\nu_s\rightarrow\nu_\tau\nu_{e(\mu)}\bar{\nu}_{e(\mu)}$& $m_s^2/768\pi^3$ & $0$\\
    $\nu_s\rightarrow\nu_\tau e^+e^-$& $(\tilde{g}_L^2+g_R^2)m_s^2/192\pi^3$ & $1$\\
    $\nu_s\rightarrow\nu_\tau \pi^0$& $f_\pi^2/32\pi\left(1-m_s^2/m_\pi^2\right)^2$ & $135$\\
    \hline
    \end{tabular}
\caption{Main decay channels up to $m_s\lesssim 200~\rm{MeV}$, for a $\nu_s$ mixed with $\nu_{\tau(\mu)}$ where $f_\pi =135~\rm{MeV}$, $\tilde{g}_L=-\frac{1}{2}+ \sin^2 \theta_W$, $g_R=\sin^2 \theta_W$, and the electron mass is neglected~\cite{Mastrototaro:2019vug,Gorbunov:2007ak}.}
\label{decayt}
\end{table}

With a little abuse of notation, we shall refer to $\nu_H$ as $\nu_s$ and to $\nu_\ell$ as $\nu_\alpha$. In the following we assume sterile neutrinos as Majorana particles, i.e. we do not distinguish neutrinos from antineutrinos.

In a recent paper~\cite{Mastrototaro:2019vug} some of us have considered this in the context of avoiding an excessive energy loss, saturating the SN 1987A luminosity bound of Eq.~\eqref{eq:lumenX}. For a mixing between $\nu_s$ and $\nu_{\mu,\tau}$ it was found there that the active-sterile neutrino mixing angle should satisfy $\sin^2 \theta_{x s} \lesssim 5 \times 10^{-7}$ (with $x= \mu, \tau$). The case of mixing with $\nu_e$ is more complicated since it would affect not only the energy loss but also the deleptonization of the SN core~\cite{Raffelt:1992bs}. Therefore, for simplicity, in the following we will consider only the mixing between $\nu_s$ and $\nu_{\mu,\tau}$.\footnote{Note, however, that the authors of Ref.~\cite{Dar:1986wb} studied the mixing between heavy neutrinos and electron neutrinos $\nu_e$ and, using the early measurements of the 511 keV gamma-ray flux~\cite{osti_7251482}, constrained the heavy-neutrino lifetime, requiring that the heavy neutrinos had to decay inside the Galaxy, with $\nu_H    \rightarrow\nu_e\,e^+\,e^-$ as the dominant decay channel.}
Since mixing with $\nu_\mu$ or $\nu_\tau$ is equivalent, for simplicity we will show bounds in terms of $|U_{\tau s}|^2$. We note, however, that other constraints that we show for comparison do depend on the species, cf.~e.g.~Ref.~\cite{Agrawal:2021dbo} for recent plots.

\begin{figure}[t]
    \centering
    \includegraphics[width=1\columnwidth]{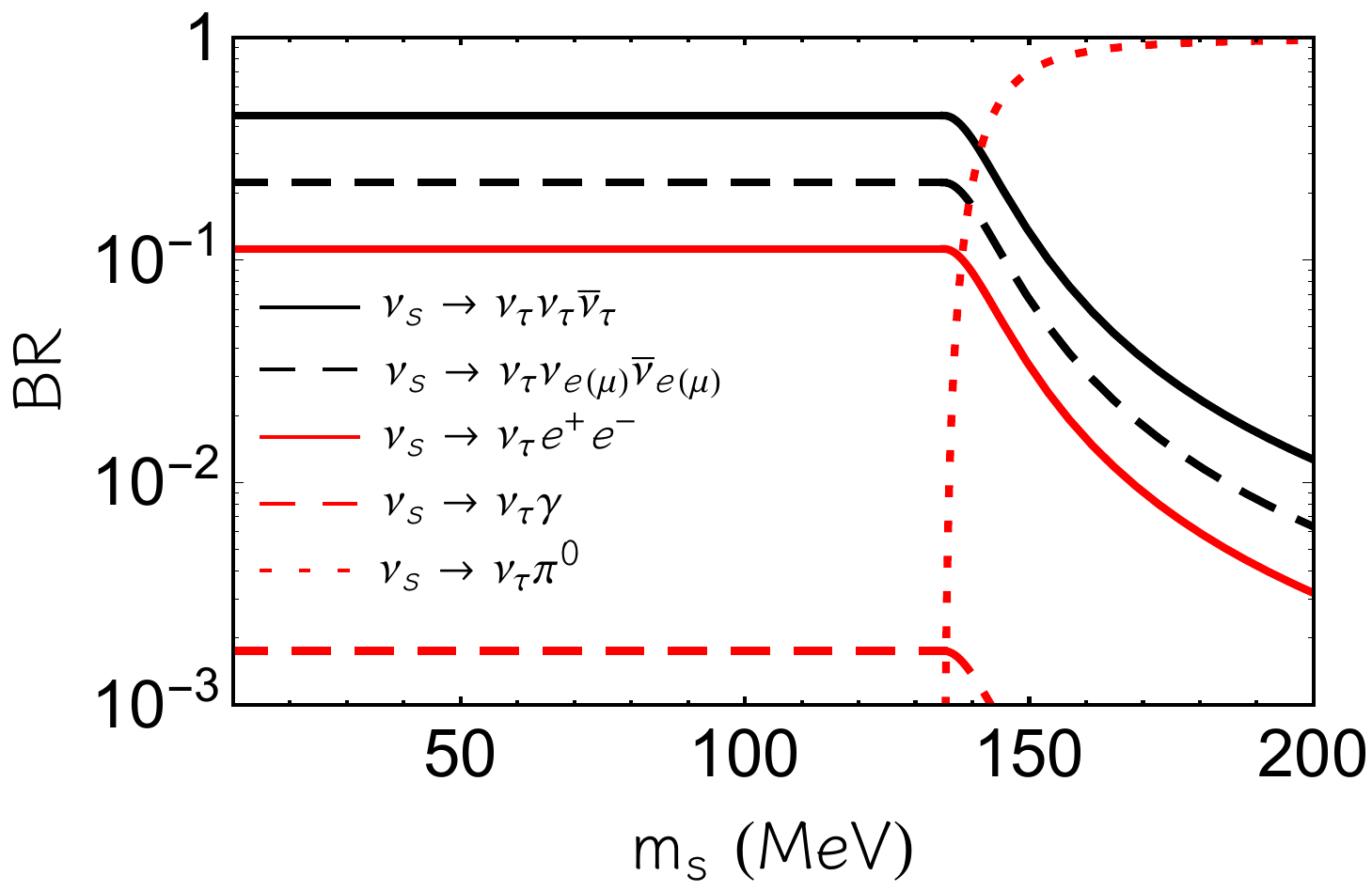}
    \caption{Branching ratios of sterile neutrinos decays in the case of mixing with $\nu_{\tau(\mu)}$, involving the processes in Table~\ref{decayt}.
    }
    \label{fig:BR_tau}
\end{figure}

The sterile neutrino distribution $f_s$ in a SN core is obtained by solving the Boltzmann equation
\begin{equation}
    \frac{\partial f_s}{\partial t}=I_{\mathrm{coll}} \,,
    \label{Boltz_eq}
\end{equation}
characterising all the possible sterile neutrino collisional interactions. In a hot core $n$, $p$, $e^\pm$, and $\nu_e$ are degenerate and the blocking factor will render pair annihilation and inelastic scattering of non-degenerate neutrinos species $\nu_{\mu,\tau}$ the dominant process for $\nu_s$ production in the case of $\nu_s$-$\nu_{\mu,\tau}$ mixing. In order to characterise these processes, we closely follow~\cite{Mastrototaro:2019vug}. Details of this calculation are recalled in Appendix \ref{app:sterile}. 

Referring to Eq.~(\ref{eq:npos}), the produced sterile energy spectrum is given by (see Ref.~\cite{Mastrototaro:2019vug})
\begin{equation}
\frac{dN_{s}^{0}}{dE_s}= \int dt \int_{\rm SN} dV \frac{pE_s}{2\pi^2}\frac{df_s}{dt} \,\ .
    \label{eq:dnps}
\end{equation}

After being produced, sterile neutrinos may decay through the channels listed in Table~\ref{decayt}. The total decay length in Eq.~(\ref{eq:npos}) is given by 
\begin{equation}
l_{s}=\frac{\gamma v}{\Gamma_{\mathrm{tot}}} \,\ ,
\label{eq:decayels}
\end{equation}
where $\Gamma_{\mathrm{tot}}$ is the sum of the decay widths $\Gamma$ of all the decay processes and $\gamma=(1-\beta^2)^{-1/2}$ is the Lorentz factor, with $\beta=p/E_s$. The total decay rate $\Gamma_{\mathrm{tot}}$ and its dependence on $m_s$ can be directly obtained by summing over the channels in Table~\ref{decayt}. In Fig.~\ref{fig:BR_tau}, we show the branching ratios of the processes in Table~\ref{decayt} for $m_s\lesssim 200~\rm{MeV}$. We note that the decay channel into electron-positron pairs has a branching ratio of $\sim 10 \%$ for masses $m_s < 135$~MeV, while its contribution becomes suppressed for higher $m_s$ when decays into $\pi^0$  become dominant.  

Plugging Eqs.~\eqref{eq:dnps} and \eqref{eq:decayels} into Eq.~\eqref{eq:npos} and determining the average number of positrons produced from the branching fractions according to Eq.~\eqref{eq:averagepositrons}, we obtain the injected positron flux from a SN.
\begin{figure}[t!]
\vspace{0.cm}
\includegraphics[width=1\columnwidth]{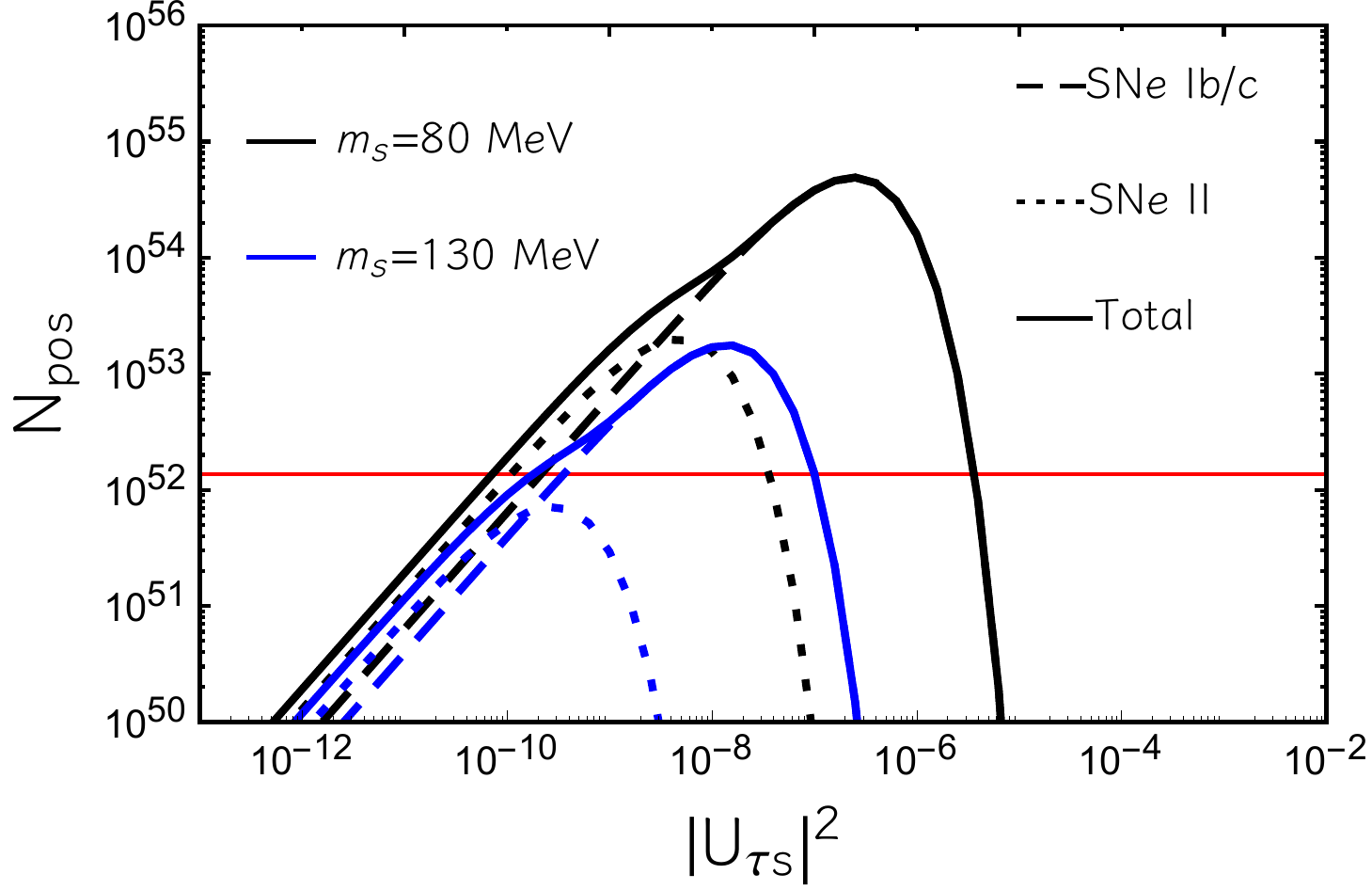}
\caption{The number of positrons injected in the Galaxy by decays of $\nu_s$ produced by SNe Ib/c (dashed lines) and SNe II (dotted lines). The continuous curves show the sum of the two contributions. The two cases are $m_{s}=80~\MeV$ (black line) and {$m_{s}=130~\MeV$} (blue line). The horizontal red line represents the bound on the number of positrons $N_{{\rm pos}}=1.4\times 10^{52}$.}
\label{fig:sterile_positrons}
\end{figure}
\begin{figure*}[t!!]
\includegraphics[width=1.54\columnwidth]{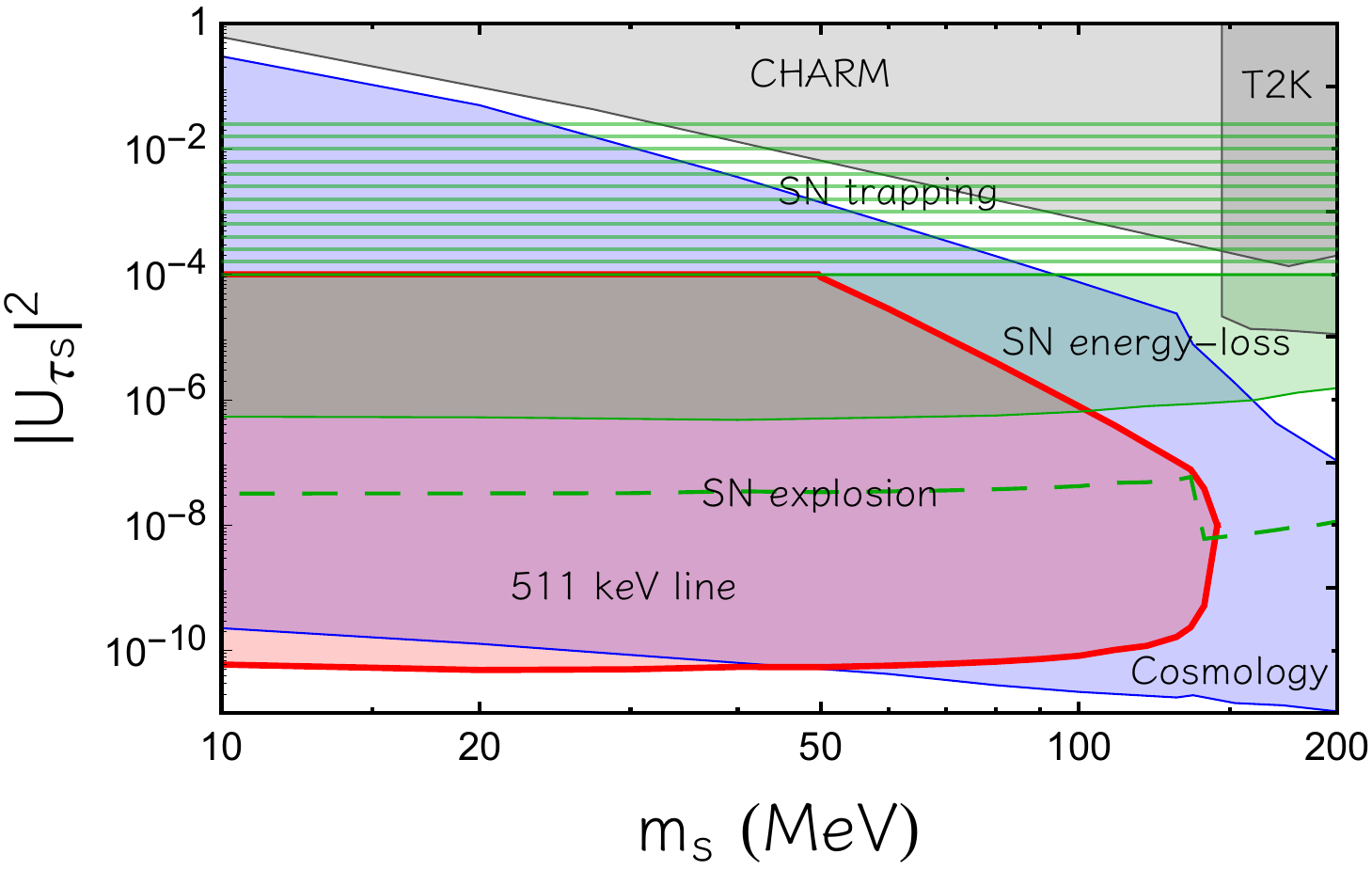}
\caption{Bounds in the plane $(m_s,|U_{\tau s}|^2)$. The bound from the 511 keV gamma-ray line is the red area. The SN 1987A energy-loss bound is the green (hatched) area in the free-streaming~\cite{Mastrototaro:2019vug} (trapping~\cite{Dolgov:2000jw}) regime. The bound from SN explosion energy~\cite{Sung:2019xie} is represented by the green dashed line. Cosmological bounds from BBN and CMB are shown as the blue area~\cite{Mastrototaro:2021wzl,Sabti:2020yrt}. We also depict the limits from laboratory experiments CHARM and T2K (grey bands)~\cite{Bolton:2019pcu}. We remark that 511 keV line limits on the $|U_{\mu s}|^2$ mixing are equivalent.
}
\label{results_confronto}
\end{figure*}
In Fig.~\ref{fig:sterile_positrons} we show the number of positrons that are produced \emph{inside} the Galaxy as a function of the mixing parameter $|U_{\tau s}|^2$ for two $\nu_s$ masses, $m_{s}=80$~MeV (black lines) and $m_{s}=130$~MeV (blue lines), respectively. We show the total number of positrons as the solid lines. Furthermore, we also indicate the individual contributions coming from SNe Ib/c (dashed lines) and from SNe II (dotted lines). The horizontal red line represents $N_{\rm pos}=1.4\times 10^{52}$, corresponding to the number of positrons saturating the bound. Referring to Eq.~(\ref{cases}), the $\nu_s$ production rate scales as $g_{X,P}^2 \sim |U_{\tau s}|^2$ and the decay rate scales as $g_{X,D}^2 \sim |U_{\tau s}|^2$. Since for the considered mixing and masses $l_X \ll r_G$, only the conditions in Eqs.~(\ref{rii}) and (\ref{ri}) are realized. For values of the mixing $|U_{\tau s}|^2 \lesssim 10^{-10}$, the exact threshold depending on $m_{s}$, the number of positrons in the Galaxy is suppressed since the $\nu_s$ production in the SN is low [according to Eq.~(\ref{rii}), it scales as $|U_{\tau s}|^2$]. Although all $\nu_s$s decay inside the Galaxy they are not abundant enough to saturate the positron bound (red line in the Figure).\\
As the mixing increases, more $\nu_s$ are produced in the SN core, but a large fraction decays before reaching the photosphere [see Eq.~(\ref{ri})]. Since the photosphere radius $r_{II}$ for Type II SNe is two orders of magnitude larger than the corresponding one $r_{I}$ for Tpye Ib/c,  the Type II SNe contribution to the positron flux is dominant for lower values of the coupling, while for larger couplings their contribution is exponentially suppressed, whereas  the SNe Ib/c contribution becomes the leading one. In the region, $10^{-10} \lesssim |U_{\tau s}|^2 \lesssim 10^{-8}$ (somewhat depending on $m_s$), the contributions of both SN types are relevant: in particular, the SNe II contribution starts to be suppressed and the SNe Ib/c one is still increasing.

\subsection{511~keV line bound}
Following the strategy outlined in Sec.~\ref{sec:positrons}, i.e. applying the constraint of Eq.~\eqref{eq:nposbound} on the positron emission of SNe, we can now constrain the sterile neutrino parameter space. The red region in Fig.~\ref{results_confronto} shows the 511~keV line bound in the $|U_{\tau s}|^2$ \emph{vs} $m_s$ plane, evaluated using as benchmark case the $18~M_\odot$ SN model and $\lambda=1$~kpc. In the small mass limit ($m_s\lesssim 50$~MeV), values of the mixing $|U_{\tau s}|^2\gtrsim 10^{-10}$ are excluded. However, we remark that the astrophysical uncertainties previously discussed may sizeably affect the bound. Indeed, as further discussed in Appendix~\ref{app:uncertainties}, since the $\nu_s$ production depends on the SN core temperature, which increases in function of the SN progenitor mass, the constraint suffers from an uncertainty of $\sim 2$ orders of magnitude from the
progenitor mass, excluding in the small mass limit $|U_{\tau s}|^2\gtrsim 10^{-9}$ with the $8.8~M_\odot$ SN model and $|U_{\tau s}|^2\gtrsim 10^{-11}$ with the $25~M_\odot$ SN model. In addition, the smearing of the 511 keV line implies another $\sim 1$ order of magnitude uncertainty when varying from a $\lambda=0$~kpc (i.e., no smearing) to $\lambda=10$~kpc (i.e.,~a quite extreme value for the smearing).

For comparison we show the laboratory bounds from the beam dump experiment CHARM and from the accelerator T2K~\cite{Bolton:2019pcu} (both in grey) as well as the SN 1987A energy-loss bound (green area)~\cite{Mastrototaro:2019vug}. We remark that this latter bound was obtained in Ref.~\cite{Mastrototaro:2019vug} considering that the  $\nu_s$ are free-streaming in the SN core. This restricts the applicability to mixing parameters $|U_{\tau s}|^2 \lesssim 10^{-4}$. For larger values of the mixing, the $\nu_s$ would be trapped in the SN core, contributing to energy transfer. Since, as far as we know, there are no recent studies of this regime, in our figure we show the exclusion region up to $|U_{\tau s}|^2 \simeq 2.5\times 10^{-2}$ (green hatched in our figure) derived in Ref.~\cite{Dolgov:2000jw}. As already mentioned, the SN 1987A energy-loss bound can be improved towards lower mixing requiring that the energy deposition in the SN envelope due to $\nu_s$ decays into electromagnetic channel does not lead to too energetic SN explosions~\cite{Sung:2019xie} [see Eq.~(\ref{eq:lumenX_1})], obtaining the limit $|U_{\tau s}|^2\lesssim 3\times 10^{-8}$ for $m_s \ll 100$~MeV and $|U_{\tau s}|^2\lesssim 6\times 10^{-9}$ for $m_s > 135$~MeV, when the decay is dominated by pion production.\footnote{The produced neutral pions subsequently decay into $2\gamma$ with a mean decay time $\tau\sim 8\times 10^{-17}~\rm{s}$.} We indicate this bound with a green dashed line, since in our opinion it should be taken only as indicative. Further substantiation would require a proper SN simulation of the effect of energy deposition (see Ref.~\cite{Rembiasz:2018lok} for a first study in this direction).

We point out that for $m_s \lesssim 100$~MeV the 511~keV signal allows us to strengthen the SN 1987A energy-loss bound and the one from energy deposition by several orders of magnitude, down to  $|U_{\tau s}|^2 \simeq 10^{-10}$. 
We stress that this constraint remains significantly more stringent than the other SN bounds even after the relaxation due to the astrophysical uncertainties discussed above are taken into account.
The bounds end at $m_s \sim 135$~MeV, since for larger masses the branching ratio for production of electron-positron pairs becomes severely suppressed, the decays being dominated by the pion production (see Fig.~\ref{fig:sterile_positrons}).

For comparison we show that these small mixing parameters can be probed only with cosmological arguments, evaluating the impact of $\nu_s$ decays on Big-Bang Nucleosynthesis (BBN) and on the effective number of neutrino species $N_{\rm eff}$ measured by the Cosmic Microwave Background (CMB)~\cite{Mastrototaro:2021wzl,Sabti:2020yrt} (blue area). However, as shown in Refs.~\cite{Gelmini:2004ah,Gelmini:2008fq}, in non-standard cosmological scenarios with low-reheating temperature, i.e., $T_{\rm RH} \ll 100$~MeV, these bounds can be easily relieved. Therefore, it is important to have an independent astrophysical bound in this region which is not affected by cosmological assumptions.

\section{Dark photons}\label{sec:darkphotons}

\subsection{Positron flux}
\label{sec:dpposflux}
The dark photon (DP) is a $U(1)^\prime$ gauge boson \emph{kinetically} mixed with the SM photon~\cite{Okun:1982xi,Holdom:1985ag}. In this context, the relevant terms in the Lagrangian containing the DP are~\cite{Holdom:1985ag,Foot:1991kb}
\begin{equation}
    \mathcal{L}=\frac{1}{2}m_{A^\prime}\,A^\prime_{\mu}\, A^{\prime\mu}-\frac{1}{4}\,F^\prime_{\mu\nu}\,F^{\prime\mu\nu}-\frac{\epsilon}{2}\,F^\prime_{\mu\nu}F^{\mu\nu}\,,
\end{equation}
where $A'$ is the DP field, $\epsilon$ the mixing parameter, $F_{\mu\nu}$ the electromagnetic field strength tensor and $F'_{\mu\nu}$ the same for the DP.

Being massive, DPs have both transverse ($T$) and longitudinal ($L$) degrees of freedom. Their production in a SN core has been recently calculated in a series of papers (see, e.g., \cite{Kazanas:2014mca,Chang:2016ntp,Hardy:2016kme,Stetina:2017ozh,DeRocco:2019njg}). In this Section we closely follow the calculation of Ref.~\cite{DeRocco:2019njg}, whose details are recalled in Appendix \ref{app:dpprod}.

The energy spectrum of the produced DPs is given by
\begin{eqnarray}
\frac{dN_{A'}^{0}}{dE}= \int dt \int_{\rm SN} dV  \frac{dN_{A'}^{0}}{dV\,dE\,dt} \,\ ,
\label{eq:dNA'}
\end{eqnarray}
where the number of DPs produced per unit volume and time takes contribution from both the $L$ and $T$ modes
\begin{equation}
    \frac{dN^{0}_{A'}}{dV dt}=
    \frac{dN^{0}_{A'}}{dV dt}\bigg|_L+\frac{dN^{0}_{A'}}{dV dt}\bigg|_T \,\ .
\end{equation}

After being produced in the SN core, the DPs can decay into $e^+ e^-$~\footnote{We mention that DPs can decay also into three photons through an electron loop, but this decay channel is relevant only for $m_{A'}<2\,m_e$~\cite{Redondo:2008ec}.} with a decay length given by
\begin{equation}
    l_{e}=\frac{\gamma v}{\Gamma_{A'\rightarrow e^+\,e^-}}\,,
\end{equation}
where
\begin{equation}
 \Gamma_{A'\rightarrow e^+\,e^-}=\frac{1}{3}\alpha \epsilon^2 m_{A^\prime} \sqrt{1-\frac{4 m_e^2}{m_{A^\prime}^2}}\left(1+\frac{2 m_e^2}{m_{A^\prime}^2}\right)    
\end{equation}
is the total decay width. Indeed, for DPs with mass below $\sim 200$ MeV this is the only available decay channel. Therefore, we can use $n_{\rm pos, A'}=1$ in Eq.~(\ref{eq:npos}).

\begin{figure}[t]
\includegraphics[width=1\columnwidth]{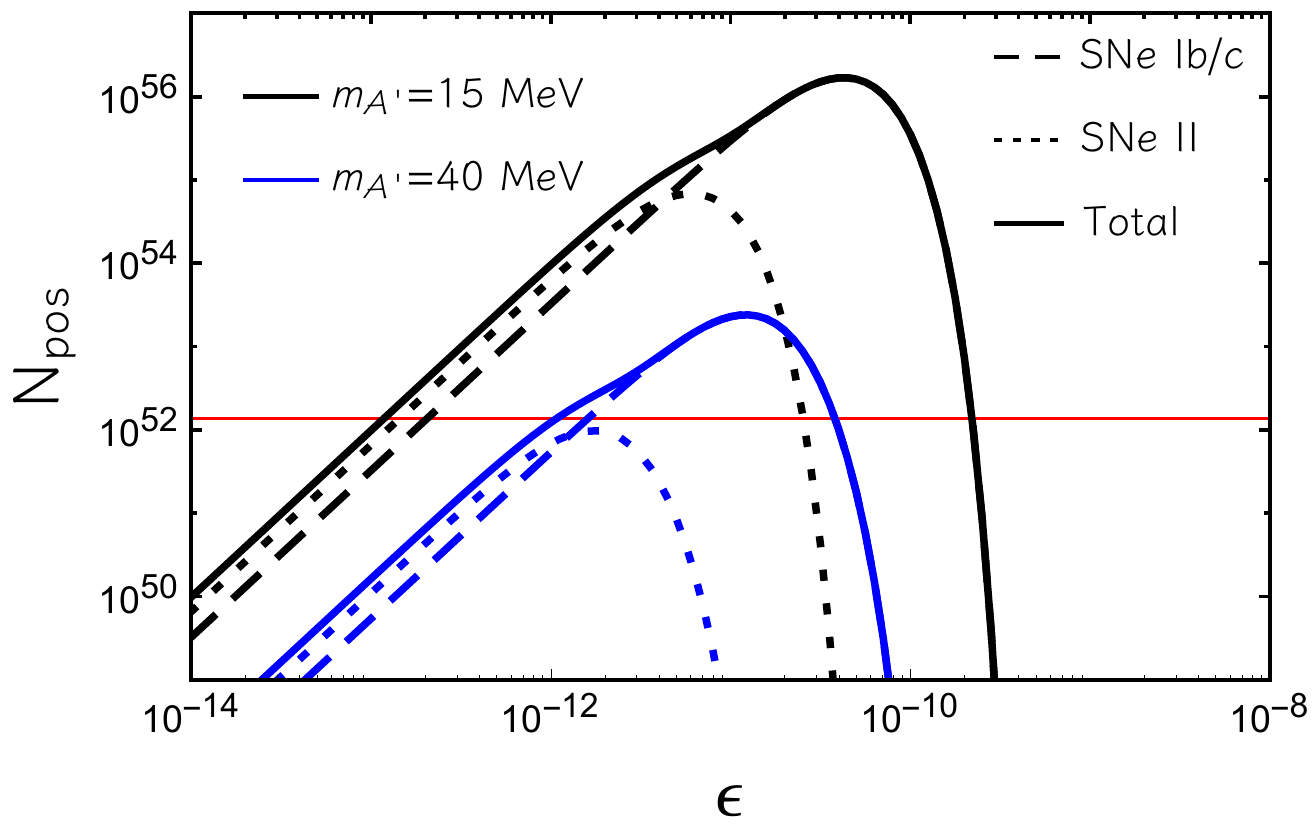}
\caption{The number of positrons injected in the Galaxy by decays of DPs produced by SNe Ib/c (dashed lines) and SNe II (dotted lines) for $m_{A'}=15~\MeV$ (black line) and $m_{A'}=40~\MeV$ (blue line). The continuous curves show the sum of the two contributions. The horizontal red line represents the bound on the number of positrons $N_{\rm pos}=1.4\times 10^{52}$.}
\label{fig:NposDP}
\end{figure}

In Fig.~\ref{fig:NposDP} we show the number of produced positrons as a function of the mixing $\epsilon$ for two DP masses $m_{A'}=15$~MeV (black lines) and $m_{A'}=40$~MeV (blue lines) using the same style of Fig.~\ref{fig:sterile_positrons} for sterile neutrinos. Notably, the general trend in the figure is similar to what shown in Fig.~\ref{fig:sterile_positrons}. Referring to Eq.~(\ref{cases}), the DP production rate scales as $g_{X,P}^2 \sim \epsilon^2$ and the decay rate scales as $g_{X,D}^2 \sim \epsilon^2$. For values of the mixing $\epsilon \ll 10^{-12}$ (depending on $m_{A'}$), the number of positrons in the Galaxy is suppressed. Indeed, the DP production in the SN scales as $\epsilon^2$ [see Eq.~(\ref{rii})]. Increasing the mixing parameter, a large fraction of DPs decay before reaching the photosphere leading to an exponential suppression of the signal for $\epsilon \gtrsim 10^{-10}$ [see Eq.~\eqref{ri}]. In an intermediate range for $10^{-12}\lesssim \epsilon \lesssim 10^{-11}$, there is  a change in the slope of the curve where both the contributions from SN Ib/c and II are relevant.
  \\

\subsection{511~keV line bound}

\begin{figure*}[t!!]
\vspace{0.cm}
\includegraphics[width=1.54\columnwidth]{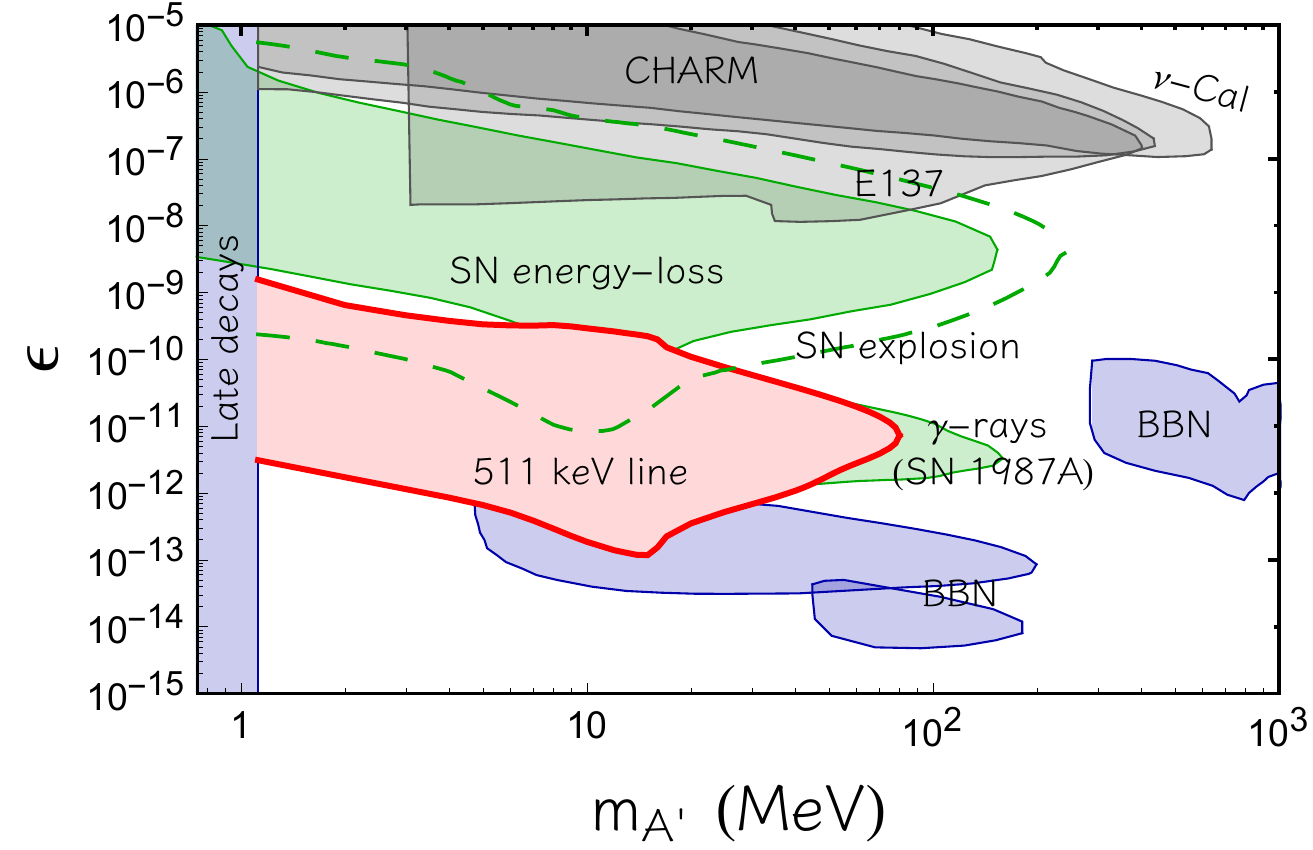}
\caption{Bounds on the DP parameter space. The red region is the bound from 511 keV gamma-ray line. In addition, we show bounds from SN 1987A~\cite{Chang:2016ntp,DeRocco:2019njg,Sung:2019xie} in green (the bound of Ref.~\cite{Sung:2019xie} is shown as a dashed line), cosmology~\cite{Fradette:2014sza,Redondo:2008ec,Li:2020roy} in blue and laboratory experiments~\cite{Gninenko:2012eq,Batell:2014mga,Blumlein:2011mv,Blumlein:2013cua} in grey, taken from Ref.~\cite{Agrawal:2021dbo}.}
\label{fig:dp_bound}
\end{figure*}

To obtain a bound on the DP parameter space we can now follow the same steps discussed in the previous Section for sterile neutrinos. Indeed, in the case of DPs the same basic argument has already been employed by De Rocco \emph{et al.} in Ref.~\cite{DeRocco:2019njg}. Before presenting our bound, we briefly note the main differences and improvements in our analysis.

The argument used in Ref.~\cite{DeRocco:2019njg} is that the SPI measurements constrain the Galactic center positron annihilation rate to be smaller than a few $\times 10^{43}$~s$^{-1}$~\cite{Prantzos:2010wi,Raffelt:1996wa}. Since electron-positron annihilation seems to be in equilibrium, one can also take the previous as a bound on the positron production rate. Assuming a Galactic SN rate of 2 events per century, in Ref.~\cite{DeRocco:2019njg} it was estimated that the previous bound would be saturated if a single SN emits more than $10^{53}$ positrons. However, this argument is somewhat oversimplified since it does not take into account the specific distribution of the positrons in the Galaxy, assuming instead that it has the same morphology of the detected 511 keV line. As we discussed in Sec.~\ref{sec:positrons}, exploiting the longitude and latitude  distributions of the 511 keV gamma-ray flux provided by an analysis of SPI data~\cite{Siegert:2019tus}, and comparing it with the expected 511 keV line signal produced by positrons from FIP decays, as traced by the core-collapse probability distribution, one gets a stringent requirement for the number of emitted positrons, i.e., a factor $\sim 7$ smaller [see Eq.~(\ref{eq:nposbound})].

Our bound from the 511 keV line is the red region shown in Fig.~\ref{fig:dp_bound} in the $\epsilon$ \emph{vs} $m_{A'}$ plane. We find only a small improvement in the upper bound ($\epsilon\sim 10^{-9}$) with respect to the constraint in Ref.~\cite{DeRocco:2019njg}, since in that region of the parameter space the number of positrons is exponentially suppressed as $\epsilon$ increases [see Eq.~\eqref{ri}], therefore a reduction in the maximum number of produced positrons, being ${\mathcal O}(1)$, has a relatively little impact on the bound. On the other hand, our improved constraining strategy has a stronger impact on the lower bound. In particular, as further discussed in Appendix~\ref{app:dpprod}, in the small mass limit ($m_{A'}\lesssim 5$~MeV), where the DP abundance is dominated by resonant $L$ modes production, the bound scales as $\sim m_{A'}^{-1}$. Then,  we exclude $\epsilon\gtrsim 6.6\times 10^{-13}$ for $m_{A'}=5$~MeV, strengthening the lower bound in Ref.~\cite{DeRocco:2019njg} by more than a factor of 2, due to the more stringent limit on the positron injection per SN. For $5\,{\rm MeV}\lesssim m_{A'} \lesssim 15$~MeV, the dominant contribution to the DP production is from the $T$ modes resonantly produced when the DP mass is of the order of the plasma frequency $\omega_{\rm pl}$, $m_{A'}\approx\omega_{\rm pl}$, the 511 keV bound scales as $\sim m_{A'}^{-2}$ and excludes $\epsilon\gtrsim 1.2\times 10^{-13}$ at $m_{A'}\approx 15$~MeV. Since in the SN core $\omega_{\rm pl}\lesssim 15$~MeV, for larger masses there is no resonant production, therefore the bound worsens. As observed in Fig.~\ref{fig:NposDP}  and commented on before, for $\epsilon \gtrsim 10^{-11}$ the contribution of SNe II becomes subleading, producing a change in the slope of the positron spectrum, reflected in the trend of the bound  at $m_{A'}\sim 50$~MeV. Our more refined constraining strategy, which implies a more stringent upper limit on the number of injected positrons per SN, allows us to enlarge the excluded region  of Ref.~\cite{DeRocco:2019njg}, probing DP masses up to $m_{A'}\approx 80$~MeV. 
Also in this case, the bound is affected by astrophysical uncertainties, as discussed in Appendix~\ref{app:uncertainties}. 
In the small mass limit ($m_{A'}\lesssim 15$~MeV) the constraint is almost independent of the SN progenitor mass, since ALPs are resonantly produced, with a weak dependence on the details of the production process and the SN model. 
For heavier dark photons there is no more resonant production and the bound strengthens as the SN progenitor mass increases, since the flux of dark photons produced off-resonance is larger due to a 
higher temperature, constraining masses up to $m_{A'}\sim 50$~MeV for the $8.8~M_\odot$ SN model and $m_{A'}\sim 110$~MeV for the $25~M_\odot$ SN model. 
In addition, the lower bound suffers roughly an order of magnitude uncertainty from the smearing of the 511 keV signal, just as in the sterile-neutrino case, so that we can exclude masses up to $\sim 95$~MeV assuming no smearing and only $m_{A'}\lesssim 40$~MeV for $\lambda=10$~kpc smearing scale.

In Fig.~\ref{fig:dp_bound} we compare our new bound from the 511~keV line with other existing bounds on the DP parameter space in the $\epsilon$ \emph{vs} $m_{A'}$ plane. Notably, these bounds are {from the energy loss of} SN 1987A ~\cite{Chang:2016ntp,DeRocco:2019njg}  (green regions), cosmology~\cite{Fradette:2014sza,Redondo:2008ec,Li:2020roy} (blue regions), and laboratory experiments~\cite{Gninenko:2012eq,Batell:2014mga,Blumlein:2011mv,Blumlein:2013cua} (grey regions).

We emphasize that astrophysics and cosmology  allow one to probe mixing parameters much smaller than those accessible in direct  laboratory experiments: the strongest experimental bound is from SLAC Beam Dump E137~\cite{Batell:2014mga}, excluding $\epsilon\gtrsim 10^{-8}$. The energy-loss bound from SN 1987A~\cite{Chang:2016ntp}  {excludes} $\epsilon\gtrsim O(10^{-9})$ for $m_{A'}\gtrsim 1~\MeV$ and down to $\epsilon\gtrsim 5\times 10^{-11}$ for masses $m_{A'}\approx 15~\MeV$.  The 511 keV bound improves the SN 1987A energy-loss bound by almost three orders of magnitude. For masses $1~\MeV \lesssim m_{A'} \lesssim 7$~MeV and mixing $\epsilon\sim O(10^{-9})$ there is a small region for $m_{A'}< 10~\MeV$ which seems to be excluded neither by SN energy-loss argument nor by the 511 keV bound. However,  this small region would be constrained by the requirement that decays occurring inside the SN envelope would not lead to excessive explosion energy~\cite{Sung:2019xie} (dashed dark green line), which allows one to exclude $\epsilon\gtrsim 2\times10^{-10}$ for $m_{A'}\gtrsim 1~\MeV$.\footnote{The apparent improvement at the upper boundary compared to Ref.~\cite{Chang:2016ntp} is a consequence of us choosing to show the most ``robust'' limit from~\cite{Chang:2016ntp} as the shaded green area. Taken by itself the tightened energy loss constraint from~\cite{Sung:2019xie} [cf. also Eq.~\eqref{eq:lumenX_1}] yields a significant improvement only on the lower boundary of the excluded region.}

Furthermore, as discussed in Refs.~\cite{DeRocco:2019njg,Kazanas:2014mca}, if a large number of DPs decay just outside the photosphere, they can create a fireball, i.e.~an optically thick plasma of electrons, positrons, and photons, which absorbs  the positrons produced in the DP decay. Therefore, the 511 keV bound is less trustable in the region where the fireball forms (namely for $m_{A'}\lesssim 20$~MeV and $\epsilon\gtrsim$ a few $\times 10^{-11}$), since we expect that a large fraction of positrons annihilate before being injected in the Galaxy. However, this region can be partially excluded imposing that the diffuse extragalactic flux of gamma rays that would be generated by these fireballs must be lower than the one measured by the Solar Maximum Mission (SMM)~\cite{doi:10.1063/1.53933} and the High Energy Astronomy Observatory (HEAO-1)~\cite{McHardy:1997fb} (see Ref.~\cite{DeRocco:2019njg} for more details). In addition, part of this region is already excluded by the SN explosion energy argument.

The 511 keV bound applies to a region of the DP parameter space with masses 1~MeV $\lesssim m_{A'}\lesssim 80$~MeV. Larger masses, up to $m_{A'}\approx 160$~MeV, can be probed through the non observation of a gamma-ray flux in coincidence with SN 1987A~\cite{DeRocco:2019njg} [the green band labelled as ``$\gamma$-rays (SN 1987A)'']. On the other hand, masses {$m_{A'} \lesssim 1$~MeV} (the blue region labelled as ``Late decays'') are excluded by assuming that {thermally produced DPs contribute to} dark matter and the photon flux from their decays on cosmological timescales does not exceed the measured intergalactic diffuse photon background~\cite{Redondo:2008ec}.

The only known way to exclude mixing parameters below the 511 keV line bound is through cosmology. Indeed, lower values of the mixing can be excluded ($\epsilon\gtrsim 10^{-14}$) by evaluating the DP impact on BBN, imposing that the light elements abundance must be not in contrast with the observed one~\cite{Fradette:2014sza}. However, as in the case of sterile neutrinos, these latter bounds can be evaded in the presence of nonstandard cosmological histories.

\section{Discussion and Conclusions}
\label{sec:conclusions}

In this paper we have investigated the physics potential of Galactic SNe to constrain MeV-ish FIPs decaying into electron-positron pairs. In particular, we focused on sterile neutrinos and DPs, produced in a SN core via mixing with ordinary neutrinos and photons, respectively. 
For suitable masses and mixings, these particles escape from the SN and decay into positrons that eventually annihilate with electrons.
Using observations of the resulting 511 keV photons by the spectrometer SPI (SPectrometer on INTEGRAL)~\cite{Siegert:2019tus}, we obtain stringent constraints for the mixing parameters. Compared to earlier studies~\cite{Dar:1986wb,DeRocco:2019njg} our main improvements lie in a more careful consideration of the signal spatial morphology (cf. Ref.~\cite{Calore:2021klc}), taking into account positron propagation~\cite{Calore:2021klc}, as well as including information from recent SN simulations with different progenitor masses (see Refs.~\cite{Fischer:2009af,Fischer:2018kdt}).
In the case of sterile neutrinos mixing with $\nu_{\mu,\tau}$, for $m_s \lesssim 100$~MeV the 511~keV signal allows one to strengthen the limit compared to the energy-loss bound by four orders of magnitude, down to $|U_{\tau s}|^2 \simeq 10^{-10}$. In the case of DPs, one excludes mixing down to $\epsilon \sim 10^{-13}$ for $m_{A'} \sim 15$~MeV. These low values of the mixing parameters are in a region not accessible by current and planned laboratory experiments~\cite{Agrawal:2021dbo}.  Only cosmological arguments can be competitive with the SN bounds on 511 keV line. However, every astrophysical or cosmological argument has its own systematic uncertainties and its own recognized or un-recognized loopholes. Therefore, to constrain FIPs it is certainly important to use as many different approaches as possible.

Furthermore, in Ref.~\cite{Sung:2019xie} it was proposed that for decaying FIPs one can improve the energy-loss bound requiring that the energy transfer from the decay products  do not lead to excessively energetic SN explosions~\cite{Sung:2019xie}. This argument would constrain the FIP luminosity going into electromagnetic channel to be two orders of magnitude lower than what derived by the energy-loss bound. The 511 keV bound on DPs provides an additional confirmation in the overlap region, but more importantly it also allows us to exclude couplings that are about two orders of magnitude smaller.
We expect this result to be general, since in order to have a contribution to the 511~keV line, one should consider FIP decays outside the SN envelope, i.e. smaller couplings with respect to the ones needed to transfer energy inside the SN.

One of the main uncertainties in our limits results from the propagation of the produced positrons (cf. Appendix~\ref{app:uncertainties}). The time-scale of the slow-down before annihilation is affected by uncertainties of up to three orders of magnitude. This leads to a corresponding variation in the number of SNe contributing to the signal. At the lower end, this number could be quite small, leading to potentially large fluctuations in the signal both from variations in the number of SNe but also in the distribution of their progenitor masses. Furthermore, the distance travelled by the positrons and therefore the length scale on which the signal is smeared compared to the original SN distribution is not well known. This too can have sizable impact on the strength of the constraints. Future theoretical and observational studies shedding light on the positron propagation would therefore be of great value.

While in our work we focused only on two  specific cases of FIPs, the same strategy can be applied to every MeV-ish particle produced in a SN core and decaying into electron-positron pairs.
In general, the take-away message is that for MeV-ish FIPs coupled with electrons, the SN bound from energy-loss can be significantly improved by different orders of magnitudes exploiting the morphology of 511~keV photon signal, reaching a region probed only by cosmological arguments.

\section*{Acknowledgments}
The work of P.C., G.L. and A.M. is partially supported by the Italian Istituto Nazionale di Fisica Nucleare (INFN) through the ``Theoretical Astroparticle Physics'' project and by the research grant number 2017W4HA7S ``NAT-NET: Neutrino and Astroparticle Theory Network'' under the program PRIN 2017 funded by the Italian Ministero dell'Universit\`a e della Ricerca (MUR).\\
The work of P.C. is partially supported by the European Research Council under Grant No. 742104 and by the Swedish Research Council (VR) under grants  2018-03641 and 2019-02337.\\
The work of L.M. is supported by the Italian Istituto Nazionale di Fisica Nucleare (INFN) through the ``QGSKY'' project and by Ministero dell'Istruzione, Universit\`a e Ricerca (MIUR).\\
The work of M.G.~is supported by a grant provided by the Fulbright U.S.~Scholar Program and by a grant from the Fundación Bancaria Ibercaja y Fundación CAI. 
M.G.~thanks the Departamento de Física Teórica and the Centro de Astropartículas y Física de Altas Energías (CAPA) of the Universidad de Zaragoza for hospitality during the completion of this work.\\
{J.J. acknowledges support from the European Union’s Horizon 2020 research and innovation programme, grant agreement No 860881-HIDDeN.}\\
{F.C.~acknowledges support from the "Agence Nationale de la Recherche”, grant n. ANR-19-CE31-0005-01.}\\
The computational work has been executed on the IT resources of the ReCaS-Bari data center, which have been made available by two projects financed by the MIUR (Italian Ministry for Education, University and Research) in the "PON Ricerca e Competitività 2007-2013" Program: ReCaS (Azione I - Interventi di rafforzamento strutturale, PONa3\_00052, Avviso 254/Ric) and PRISMA (Asse II - Sostegno all'innovazione, PON04a2A).

\appendix
\section{Sterile neutrino production}
\label{app:sterile}

The equation governing the sterile neutrinos production is the Boltzmann
equation
\begin{equation}
    \frac{\partial f_s}{\partial t}=I_{\mathrm{coll}} \,,
    \label{Boltz_eq_app}
\end{equation}
where $f_s$ is the sterile neutrino distribution and $I_{\mathrm{coll}}$ is the integral over all possible collision processes that create sterile neutrinos. It is possible to write the $I_{\mathrm{coll}}$ as
\begin{equation}
\begin{split}
    I_{\mathrm{coll}}=&\frac{1}{2E_s}\int\frac{d^3p_2}{(2\pi)^3}\frac{d^3p_3}{(2\pi)^3}\frac{d^3p_4}{(2\pi)^3}|M|^2_{s2\rightarrow34}F(f_s,f_2,f_3,f_4)\\&
    \qquad\qquad\qquad\qquad\qquad\quad\delta^4(p_s+p_2-p_3-p_4) \,\ ,
\end{split}
\end{equation}
where $|M|^2_{s2\rightarrow34}$ is the sum of the
squared amplitudes for collisional processes relevant for
the $\nu_s$ production.
{In a hot core $n$, $p$, $e^\pm$, and $\nu_e$ are degenerate, and the blocking factor will render pair annihilation and inelastic scattering of non degenerate neutrinos species $\nu_{\mu,\tau}$ as the dominant process for $\nu_s$ production in the case of $\nu_s$-$\nu_{\mu,\tau}$ mixing. The relevant
matrix elements are reported 
in Table~2 of Ref.~\cite{Mastrototaro:2019vug}.
}
 
 Then,  
\begin{equation}
    F(f_s,f_2,f_3,f_4)=f_3f_4(1-f_s)(1-f_2)-f_sf_2(1-f_3)(1-f_4) 
\end{equation}
is the usual phase space factor, including Pauli-blocking. 


Once produced, most sterile neutrinos escape freely from the SN core. Thus we assume $f_s=0$ in solving the collisional integral. Moreover $\nu_{\mu}$ and $\nu_{\tau}$ are in
thermal equilibrium in the SN core and we can describe their distribution using a Fermi-Dirac distribution:
\begin{equation}
    f_{\nu_{\mu,\tau}}=\frac{1}{e^{E/T}+1} \,\ .
\end{equation}

\section{Dark photon production}
\label{app:dpprod}

The number of dark photons produced per unit volume and time in a SN is given by~\cite{Redondo:2008aa,An:2013yfc,Redondo:2013lna,Redondo:2015iea,DeRocco:2019njg,Chang:2016ntp} (we essentially follow Ref.~\cite{DeRocco:2019njg})
\begin{equation}
\begin{split}
    \frac{dN^{0}_{A'}}{dV dt}&=
    \frac{dN^{0}_{A'}}{dV dt}\bigg|_L+\frac{dN^{0}_{A'}}{dV dt}\bigg|_T =
    \\
    &= \int \frac{d E E^2 v}{2\,\pi^2}\,e^{-E/T}(\Gamma'_{\rm abs,L}+2 \Gamma'_{\rm abs,T}) \,,
\end{split}
\end{equation}
where $v$ is the velocity, $\Gamma'_{\rm abs, L/T}$ is the absorptive width of the dark photon for the longitudinal and transverse modes {and we denoted the separate contribution of longitudinal ($L$) and transverse modes ($T$).} 

The dominant absorption process in the SN core is inverse bremsstrahlung (ibr), thus the absorptive widths are given by
\begin{eqnarray}
    \Gamma'_{\rm ibr, L/T}& =&\frac{32}{3\pi}\frac{\alpha (\epsilon_m)^2_{\rm L/T} n_n n_p}{E^3} \left(\frac{\pi T}{m_N}\right)^{3/2}\nonumber \\ &\times &  \langle\sigma_{np}^{(2)}(T) \rangle\,\left(\frac{m_{A'}^{2}}{E^2}\right)_{\rm L}\,,
\end{eqnarray}
where $n_n$ and $n_p$ are the neutron and the proton number density, $m_N=938$~MeV, $\langle\sigma_{np}^{(2)}(T) \rangle$ the averaged neutron-proton dipole scattering cross section from Ref.~\cite{Rrapaj:2015wgs}, $(\epsilon_m)^2_{\rm L/T}$ the in-medium mixing angle, and the final term is denoted with a subscript L to indicate that it is included only for the longitudinal mode. \\
In medium the mixing parameter $\epsilon$ is modified by plasma effects, therefore
\begin{equation}
    (\epsilon_m)^2_{\rm L/T}=\frac{\epsilon}{(1-{\rm Re}\Pi_{\rm L/T}/m_{A'}^{2})^2+({\rm Im}\,\Pi_{\rm L/T}/m_{A'}^{2})^2}\,,
\end{equation}
with $\Pi$ the photon polarization tensor. The real part of the polarization tensor for the two modes is given by
\begin{equation}
\begin{split}
    {\rm Re}\,\Pi_{\rm L}&=\frac{3\omega^2_p}{v^2}\,(1-v^2)\left[\frac{1}{2v}\ln\left(\frac{1+v}{1-v}\right)-1\right]\,,\\
    {\rm Re}\,\Pi_{\rm T}&=\frac{3\omega^2_p}{2v^2}\,(1-v^2)\left[1-\frac{1-v^2}{2v}\ln\left(\frac{1+v}{1-v}\right)\right]\,,
\end{split}
\label{eq:pi}
\end{equation}
with $v=k/E$ and $\omega_p$ as the plasma frequency, which in the degenerate SN plasma is given by
\begin{equation}
    \omega_p^2=\frac{4\pi \alpha_{\rm EM} n_e}{\sqrt{m_e^2+(3\pi^2\,n_e)^{2/3}}}\,,
\end{equation}
where $n_e$ is the electron number density.\\
Since in the SN core   photons are in thermal equilibrium, the imaginary part of the polarization tensor becomes
\begin{equation}
    {\rm Im}\,\Pi_{\rm L/T}=-E (1-e^{-E/T})\Gamma_{\rm abs, L/T}\,,
\end{equation}
where $\Gamma_{\rm abs, L/T}$ is the absorptive width of SM photons
\begin{equation}
    \Gamma_{\rm ibr, L/T}=\frac{32}{3\pi}\frac{\alpha n_n n_p }{E^3} \left(\frac{\pi T}{m_N}\right)^{3/2}\langle\sigma_{np}^{2}(T) \rangle\,\left(\frac{m_{A'}^{2}}{E^2}\right)_{\rm L}\,.
\end{equation}
Thus $\Gamma'_{\rm ibr, L/T}=(\epsilon_m)^2_{\rm L/T} \Gamma_{\rm ibr, L/T}$.

\begin{figure}[t]
\includegraphics[width=1\columnwidth]{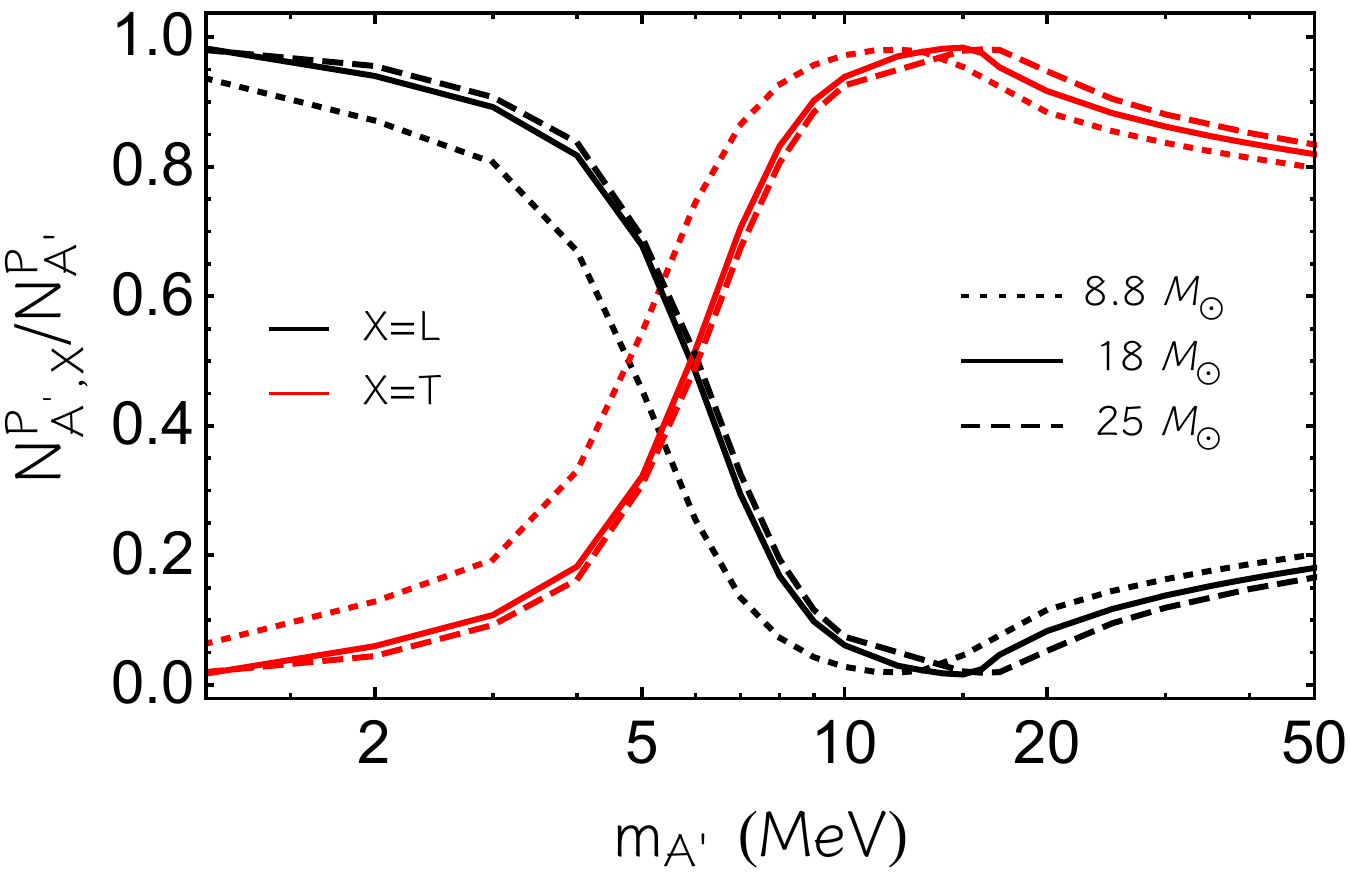}
\caption{Contribution of Longitudinal ($L$, black) and Transverse ($T$, red)  modes to DP production  in the three SN models used.
}
\label{fig:NxNtmodel}
\end{figure}

The energy spectrum of the produced dark photons is given by
\begin{eqnarray}
\frac{dN_{A'}^{0}}{dE}=&& \int dt \int_{\rm SN} dV  \frac{dN_{A'}^{0}}{dV\,dE\,dt}
\label{eq:dNA'_app}
\end{eqnarray}
where the time integral is extended over 10~s. Integrating Eq.~\eqref{eq:dNA'_app} over the energy, we obtain the number of produced dark photons $N_{A'}^{P}$. {From Eq.~\eqref{eq:pi}, when ${\rm Im}\,\Pi_{\rm L/T}\ll {\rm Re}\,\Pi_{\rm L/T}$ (condition satisfied throughout the SN core) and ${\rm Re}\,\Pi_{\rm L/T}=m_{A'}$, there is a \emph{resonance} in $\epsilon_m$. As further discussed in Refs.~\cite{An:2013yfc,Chang:2016ntp}, on resonance the number of longitudinal modes $N_{A',L}^{P}$ scales as $m_{A'}^2\epsilon^{2}$, while the number of transverse modes $N_{A',T}^{P}$ scales as $m_{A'}^4\epsilon^{2}$, without depending on the details of the production process. \\ In Fig.~\ref{fig:NxNtmodel} we show the contribution of longitudinal $N^{P}_{A',L}$ (black lines) and transverse modes $N^{P}_{A',T}$ (red lines) to the dark photon production for different SN models, namely with progenitor masses $18~M_\odot$ (solid), $25~M_\odot$  (dashed) and $8.8~M_\odot$ (dotted). The longitudinal contribution is dominant for lower masses $m_{A'}\lesssim 5$~MeV, while for larger masses the transverse modes are dominant. In particular, the peak in the dark photon production at $m_{A'}\approx 15$~MeV is related to the resonant production of the transverse modes for $m_{A'}\approx \omega_{p}$. Its location is therefore determined by the fact that in the SN core $\omega_p\lesssim 15$~MeV in the first 10~s after the SN explosion, when dark photons are produced. Indeed, the peak is shifted towards slightly larger masses as the progenitor mass increases, since the plasma frequency in the SN core becomes larger. For larger masses $m_{A'}\gtrsim 15$~MeV, DPs can be produced only off-resonance.}

\section{{Uncertainties}}\label{app:uncertainties}
In this Appendix we investigate 
the uncertainties affecting the bounds obtained in this work. As discussed in Ref.~\cite{Calore:2021klc} in the case of ALPs, one of the major sources of uncertainties is the dependence of the FIP flux on the SN model.
{As already mentioned in the main text, the number of SNe contributing to the signal at the present time may be quite modest and the composition of progenitor masses in the signal may therefore fluctuate. To estimate the impact of this we consider look at the change in the bound choosing, besides} the 18~$M_\odot$ case, two other SN models,  with progenitor mass 8.8~$M_\odot$ and 25~$M_\odot$ respectively, reported in Ref.~\cite{Fischer:2009af}.

In addition, the bound is strongly affected by the choice of the smearing scale $\lambda$ to take into account the FIP decay length and the positron propagation. Indeed, the produced 511~keV photon flux can be
smeared out due to
a non-negligible FIP decay length $l_X$ and the distance travelled by the positrons before being stopped. In order to take into account this effect, first we fixed in Eq.~\eqref{eq:npos} the Galactic radius $r_{\rm G}=1$~kpc, being  the smallest extent of the Galaxy, i.e. the vertical direction. Second, in order to account for the more diffuse emission,
we smeared the SN distribution over a scale $\lambda$ to take into account the FIP decay length and the positron propagation in the Galaxy. As further discussed in Refs.~\cite{Mirizzi:2006xx,Calore:2021klc}, the SN volume distribution is given by
\begin{equation}
n_{cc} = \sigma_{cc}(r) R_{cc}(z) \,\ ,
\label{eq:ncc}
\end{equation}
where $\sigma_{cc}(r) \propto r^\zeta e^{-r/u}$, with $\zeta =4$ and $u =1.25$ kpc~\cite{Mirizzi:2006xx}, is the Galactic surface density of core-collapse events, normalized as $2\pi\int_0^\infty dr\,r\,\sigma_{cc}(r)=1$, while $R_{cc} (z)\propto 0.79\, e^{-(z/212\, {\rm pc})^2} + 0.21\, e^{-(z/636\, {\rm pc})^2}$ is the vertical distribution, approximated as a superposition of two Gaussian
distributions with different scale height for the thin and
thick disk, normalized such that $\int d\Omega\, ds\, s^2\, n_{cc} = 1$. The smearing of the SN distribution over the scale $\lambda$ is implemented by using the smeared distributions
\begin{equation}\label{eq:smearing}
\begin{split}
 \sigma'_{cc}(r)&=A\,\int_0^{\infty} ds \sigma_{cc}(s)\,e^{-|s-r|/\lambda}\,,\\
R'_{cc}(z)&=B\,\int_{-\infty}^{\infty} ds R_{cc}(s)\,e^{-|s-z|/\lambda}\,.
\\
\end{split}
\end{equation}
Here, the normalization constants $A$ and $B$ are obtained by imposing $2\pi\int_0^\infty dr\,r\,\sigma_{cc}'(r)=1$ and $\int d\Omega\, ds\, s^2\, \sigma_{cc}'\,R_{cc}'=1$.

\begin{table}[t]
    \centering
    \begin{tabular}{|c|c|c|}
    \hline
    $\lambda$ (kpc) & $\epsilon_I$ & $N_{\rm pos}$\\
    \hline
    $0$ & $\,\ \,\ \,\ 0.33 \,\ \,\ \,\  $ & $\,\ 6.8\times 10^{51} \,\ $\\
    $0.5$ & $0.33$ & $9.7\times 10^{51}$\\
    $1$ & $0.33$ & $1.4\times 10^{52}$\\
    $5$ & $0.33$ & $7.6\times 10^{52}$\\
    $10$ & $0.33$ & $2.2\times 10^{53}$\\
    $1$ & $0.24$ & $1.3\times 10^{52}$\\
    $1$ & $0.43$ & $1.5\times 10^{52}$\\
    \hline
    \end{tabular}
 \caption{Bound on the number of positrons $N_{\rm pos}$ produced from each SN for different values of the smearing scale $\lambda$ and the fraction of SNe Ib/c $\epsilon_{I}$.\\ 
 \label{smeartab}
  }
\end{table}

\begin{figure*}[t!!!]
\vspace{0.cm}
\includegraphics[width=0.868\columnwidth]{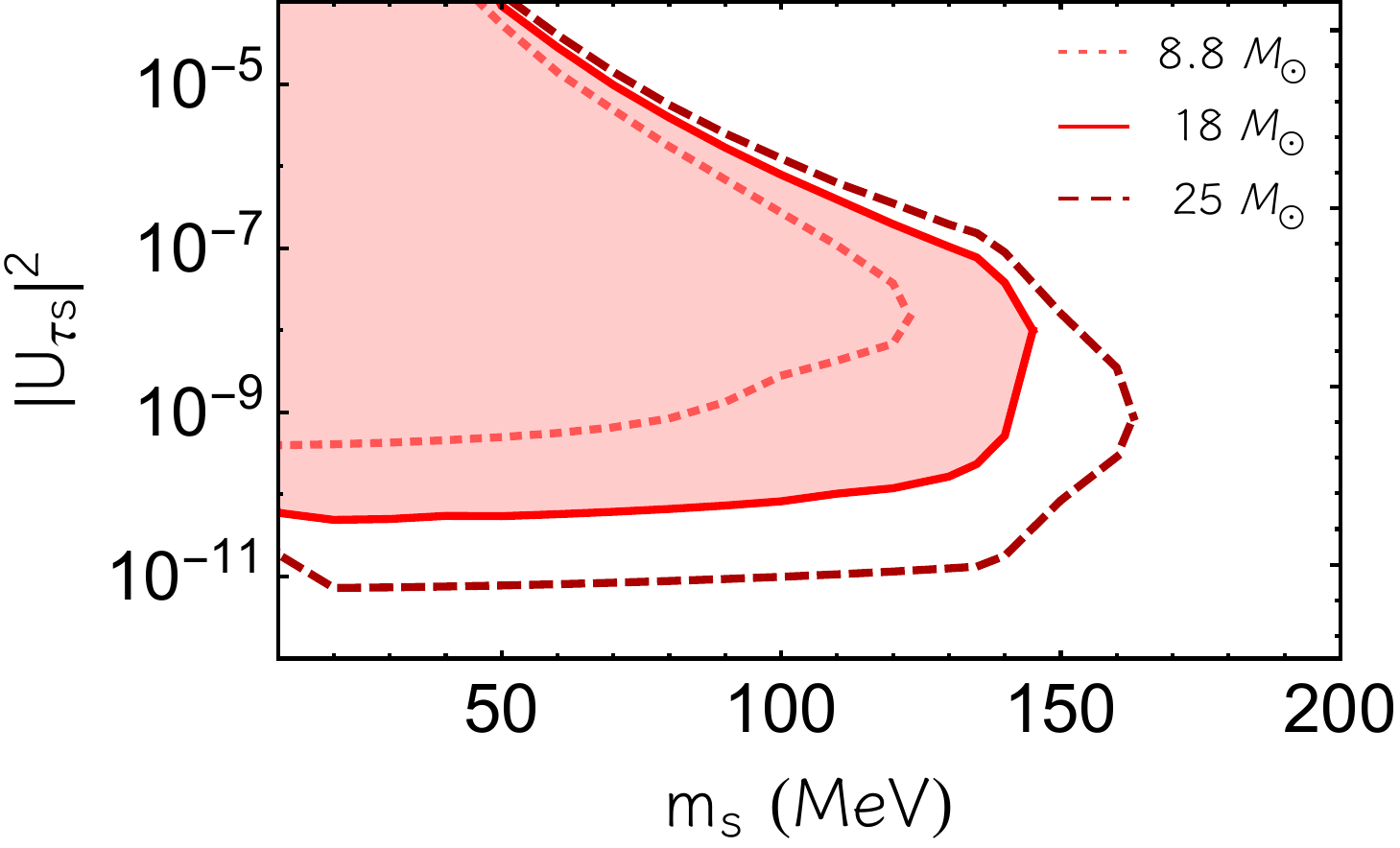}
\hspace{1.cm}
\includegraphics[width=0.85\columnwidth]{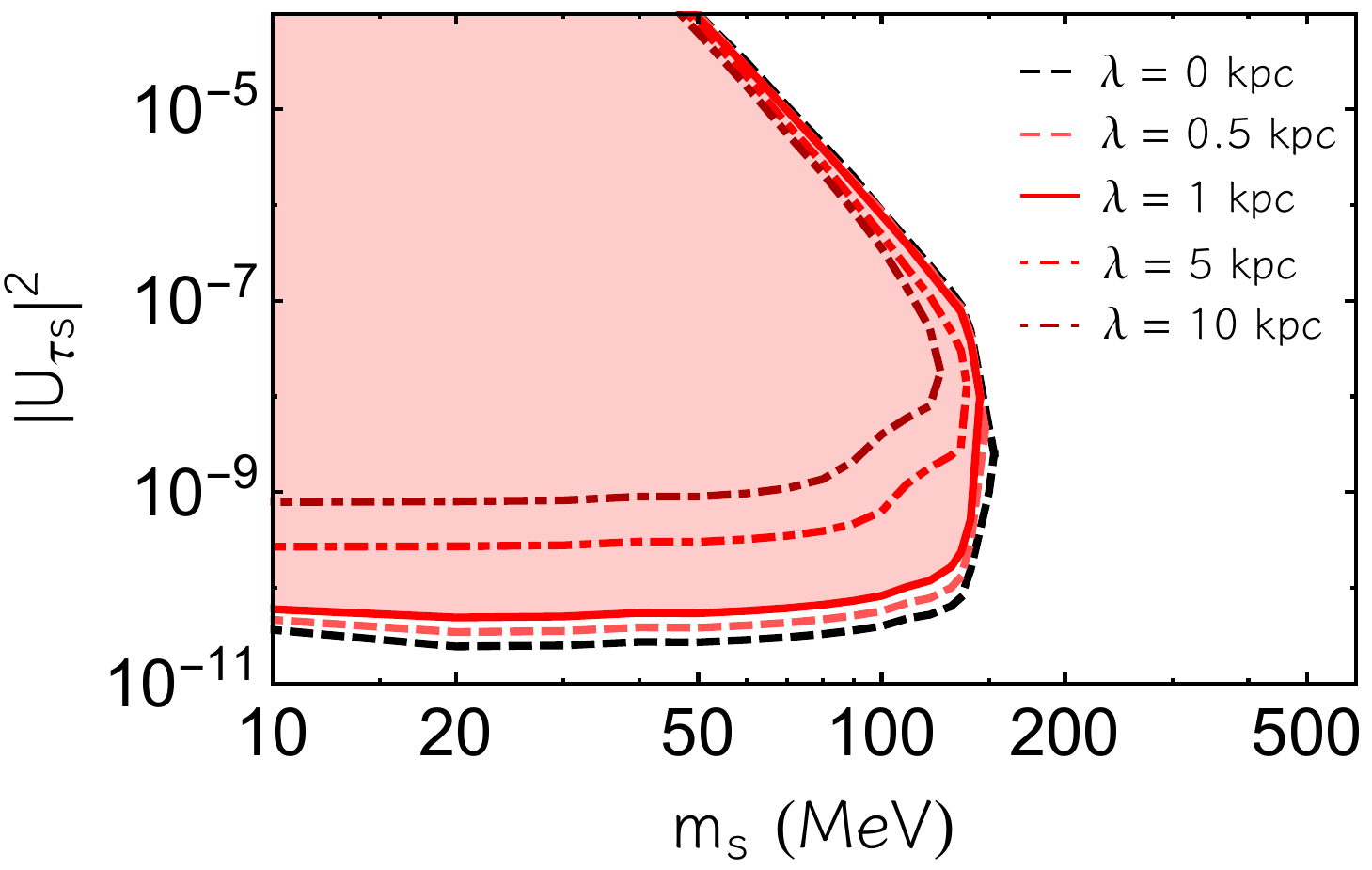}
\caption{Dependence of the 511 keV line bound for $\nu_s$ on the SN progenitor mass (\emph{left panel}) and on the smearing parameter $\lambda$ (\emph{right panel}).}
\label{fig:sterile_bound_uncert}
\end{figure*}

\begin{figure*}[t!!!]
\includegraphics[width=0.85\columnwidth]{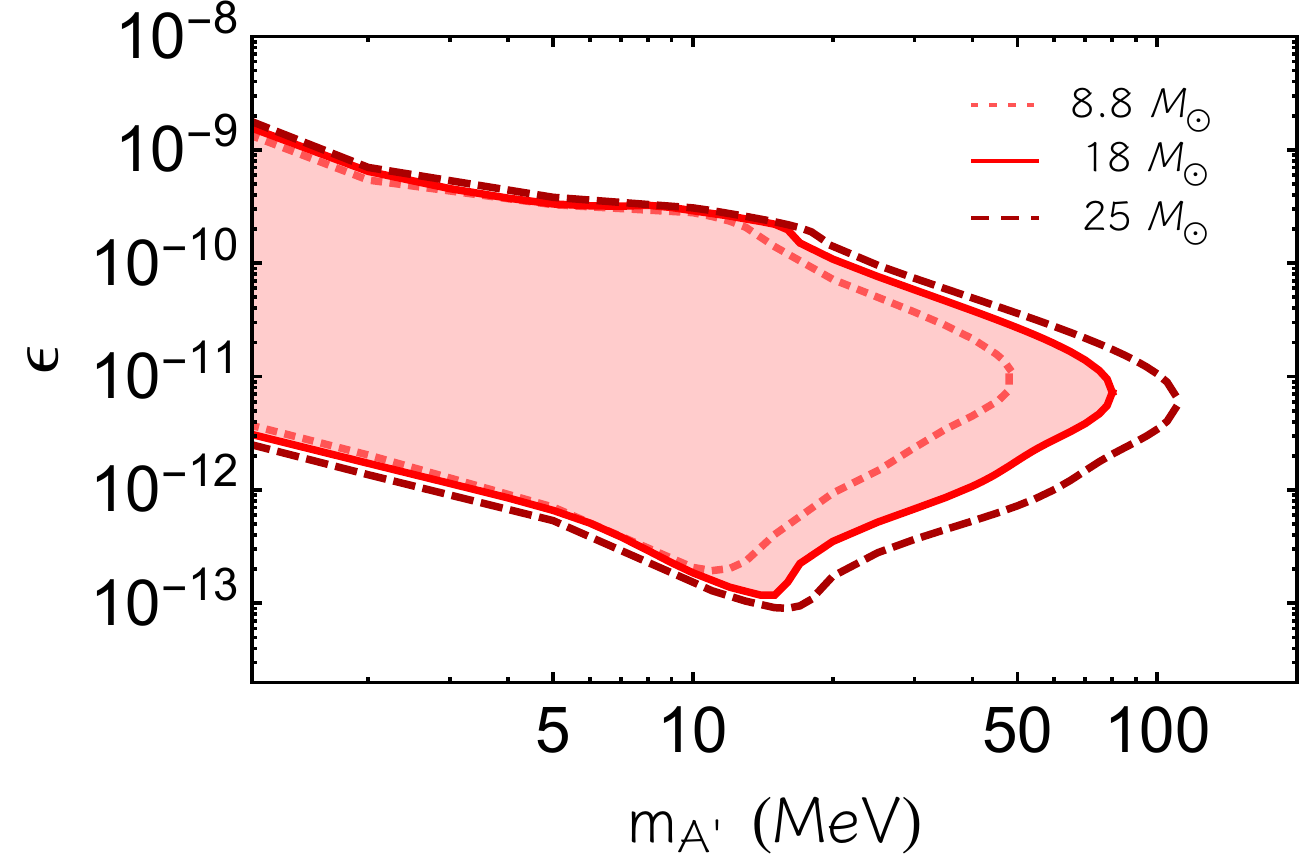}
\hspace{1.cm}
\includegraphics[width=0.85\columnwidth]{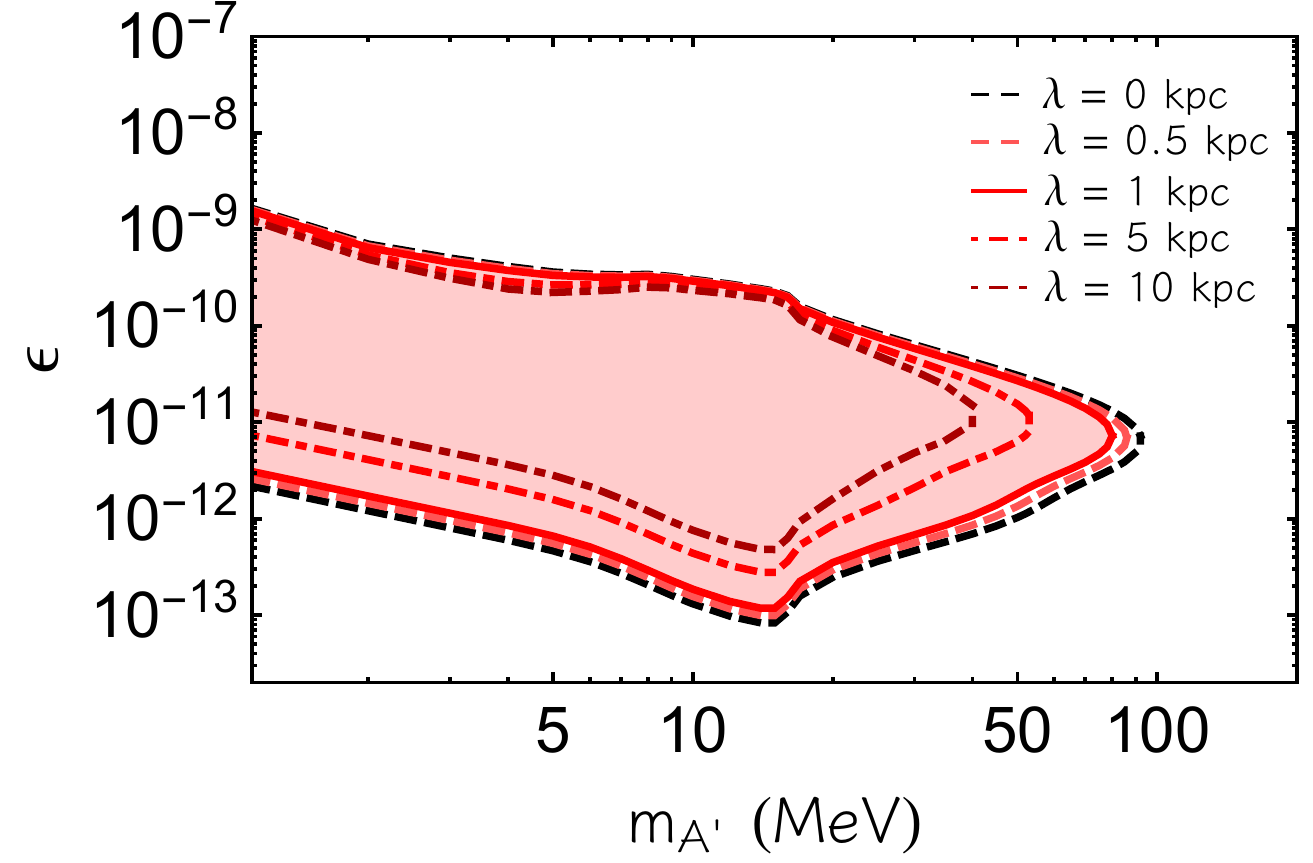}
\caption{Dependence of the 511 keV line bound for DPs on the SN progenitor mass (\emph{left panel}) and on the smearing parameter $\lambda$ (\emph{right panel}).}
\label{fig:dp_bound_uncert}
\end{figure*}

As shown in Sec.~\ref{sec:sterposflux} and Sec.~\ref{sec:dpposflux}, for the cases considered in this work $l_X\ll r_{\rm G}$, therefore the smearing scale is connected to the positron propagation length. 
The propagation conditions of low-energy positrons are poorly known and little constrained by soft gamma rays. Recently, Ref.~\cite{Siegert:2021trw} attempted to measure the positron propagation length in the context of interpreting the 511 keV line signal with old stellar distributions in the Galactic bulge, estimating it to be a few hundreds pc. Given the uncertainties at play, we study the effect of varying parametrically $\lambda$.
For this reason, in order to evaluate its impact on the 511 keV bound, we re-evaluated the constraint for different values of $\lambda$, ranging from $\lambda=0$~kpc (i.e., no smearing) to a more conservative $\lambda=10$~kpc, choosing as benchmark value $\lambda=1$~kpc, based on injection positron energies $50-100$~MeV and typical interstellar medium conditions (see, e.g. Ref.~\cite{Jean:2005af}). For larger values of $\lambda$, the photon signal becomes more featureless and, as shown in Table~\ref{smeartab}, the equivalent bound
on the number of produced positrons  from each SN tends to increase, passing from $N_{\rm pos}\lesssim6.8\times 10^{51}$ for $\lambda=0$~kpc to $N_{\rm pos}\lesssim2.2 \times 10^{53}$ for $\lambda=10$~kpc. Therefore, the exclusion region becomes smaller as the smearing scale increases (cf. the right panels of Figs.~\ref{fig:sterile_bound_uncert} and \ref{fig:dp_bound_uncert}). 

Finally, another source of uncertainty is the error on the Galactic core-collapse SN rate $\Gamma_{cc}=2.30\pm 0.48$ SNe/century, given by the sum of SNe II $\Gamma_{\rm II}=1.54\pm 0.32$~SNe/century and SNe Ib/c $\Gamma_{\rm I}=0.76\pm0.16$~SNe/century. In order to evaluate its possible impact, the bound was evaluated varying the fraction of SNe Ib/c from $\epsilon_{I}=0.6/2.46\approx 0.24$, corresponding to $N_{\rm pos}\lesssim 1.3\times 10^{52}$, to $\epsilon_{I}=0.92/2.14\approx 0.43$, i.e. $N_{\rm pos}\lesssim 1.5\times 10^{52}$, as shown in Table~\ref{smeartab}.


\subsection{Sterile neutrinos}
In {the left panel of Fig.~\ref{fig:sterile_bound_uncert}} we show the effects
of the three SN progenitor masses considered
on the $\nu_s$ bound. 
We observe that the lower part of the excluded region  is significantly dependent  on the progenitor mass.
Indeed, the bound  strengthens by two orders of magnitudes passing from 8.8~$M_\odot$ to
25~$M_\odot$ progenitors.
Indeed, the SN core temperature increases in function of the SN progenitor mass, leading to 
a larger $\nu_s$ production. Moreover, for small mixing parameters all $\nu_s$'s decay outside the SN envelope [see Eq.~(\ref{rii})].
These two circumstances would allow us to exclude a larger region of the parameter space. Conversely,  the upper part of the  bound is less affected, being the photon flux
  exponentially suppressed, as shown in Eq.~(\ref{ri}).
  
 Finally, for 25~$M_\odot$ model and $m_s\lesssim 20~\mathrm{MeV}$, the bound is reduced with different behaviour respect to 8.8~$M_\odot$ and 25~$M_\odot$ models. This effect results from constraining small $|U_{\tau s}|^2$ values for which $l_s>r_G$, causing a reduction in the electron-positron flux due to the decay of the sterile neutrinos outside the Galaxy.  \\
In the right panel of Fig.~\ref{fig:sterile_bound_uncert}, we
show the dependence  of the bound on the 
variability of the smearing scale $\lambda$. In particular, we
compare our fiducial bound with $\lambda=1~\mathrm{kpc}$ (solid red line) with bounds obtained with different smearing scales.  
As in the case of the progenitor mass, only the lower bound is significantly affected: the bound strengthens by two order of magnitudes passing from $\lambda=0$ to
$\lambda=10~\rm{kpc}$. 

Finally, we comment that the uncertainty on the Galactic SN rate has small effects on the bound. Indeed, if $\epsilon_I$ is reduced (increased) the lower bound is strengthened (relaxed) by a factor $\sim 7\%$, due to the variation in $N_{\rm pos}$ shown in Table~\ref{smeartab}.

\subsection{Dark photons}

In the left panel of Fig.~\ref{fig:dp_bound_uncert} we show the impact on the SN progenitor mass on the DP bound. Remarkably, in the small mass limit ($m_{A'}\lesssim \omega_{\rm pl,max}\approx 15$~MeV, where $\omega_{\rm pl,max}$ is the larger value of the plasma frequency in the SN core) the bounded region is rather  insensitive of the progenitor mass. Indeed, as discussed in Appendix~\ref{app:dpprod}, for these values of the mass, a \emph{resonant} production is possible and the bound scales as $\sim m_{A'}^{-1}$ when the $L$ modes are dominant ($m_{A'}\lesssim 5$~MeV) and as $\sim m_{A'}^{-2}$ when the $T$ modes are dominant ($5~\MeV \lesssim m_{A'}\lesssim \omega_{\rm pl,max}$), with a weak dependence on the details of the production process and the SN model. On the other hand, for larger masses $m_{A'}\gtrsim\omega_{\rm pl,max}$, the bound is significantly affected by the progenitor mass. In particular, the bump in the lower limit (placed at $m_{A'}\approx\omega_{\rm pl,max}$) is shifted towards lower values of the mass as the the progenitor mass decreases, since the plasma frequency becomes smaller in the SN core ($\omega_{\rm pl,max} \approx 11$~MeV in the 8.8~$M_\odot$ SN model), while we can probe heavier DPs as the progenitor mass increases, since the flux of dark photons produced off-resonance is larger due to a larger temperature.\\
In the right panel of Fig.~\ref{fig:dp_bound_uncert} we compare our fiducial bound with $\lambda = 1$~kpc (the shaded red area in the solid red line) with bounds obtained with different smearing scales. It is apparent that the shape of the bound is independent of the smearing scale, but as $\lambda$ increases, a smaller area of the DP parameter space can be excluded. Indeed, in the extreme case in which we completely neglect the positron propagation (i.e. $\lambda = 0$~kpc), DP masses up to $\sim 95$~MeV can be excluded, while for $\lambda=10$~kpc only $m_{A'}\lesssim 40$~MeV can be constrained, due to the less stringent bound on $N_{\rm pos}$ as $\lambda$ increases.

Finally, we comment that the error on the Galactic SN rate has a really small impact on the bound. Indeed, the only effect is that if $\epsilon_I=0.24$ ($\epsilon_I=0.43$), the lower bound is strengthened (relaxed) by a factor $\sim 3.5\%$, due to the different value of $N_{\rm pos}$, while the upper limit is even less modified since the number of positrons is exponentially suppressed. This uncertainty has also small impact on the largest mass that can be probed, $m_{A'}\approx 75$~MeV for $\epsilon_I=0.24$ and $m_{A'}\approx 85$~MeV for $\epsilon_I=0.43$.

\bibliographystyle{utphys}
\bibliography{references.bib}

\end{document}